\documentclass{aa}  

\usepackage{graphicx}
\usepackage{txfonts}
\begin{document}

   \title{Mysterious eclipses in the light curve
of KIC8462852: a possible explanation}

%   \subtitle{I. Overviewing the $\kappa$-mechanism}

   \author{
L. Neslu\v{s}an \inst{1}
          \and
J. Budaj \inst{1}
          }

   \institute{Astronomical Institute, Slovak Academy of Sciences,
05960 Tatransk\'{a} Lomnica, Slovak Republic\\
\email{ne@ta3.sk, budaj@ta3.sk}
             }

   \date{Received July 20, 2016; accepted ???? ??, ????}

% \abstract{}{}{}{}{} 
% 5 {} token are mandatory
 
  \abstract
{Apart from thousands of `regular' exoplanet candidates, Kepler
satellite has discovered a few stars exhibiting peculiar
eclipse-like events. They are most probably caused by disintegrating
bodies transiting in front of the star. However, the nature of the
bodies and obscuration events, such as those observed in KIC8462852,
remain mysterious. A swarm of comets or artificial alien mega-structures
have been proposed as an explanation for the latter object.}
{We explore the possibility that such eclipses are caused by the dust
clouds associated with massive parent bodies orbiting the host star.}
{We assumed a massive object and a simple model of the dust cloud
surrounding the object. Then, we used the numerical integration to 
simulate the evolution of the cloud, its parent body, and resulting
light-curves as they orbit and transit the star.}
{We found that it is possible to reproduce the basic features in the
light-curve of KIC8462852 with only four objects enshrouded in dust
clouds. The fact that they are all on similar orbits and that such
models require only a handful of free parameters provides additional
support for this hypothesis.}
{This model provides an alternative to the comet scenario. With such
physical models at hand, at present, there is no need to invoke alien
mega-structures for an explanation of these light-curves.}

% -- (Stars): planetary systems -- (ISM:) dust, extinction}
% Celestial mechanics --

   \maketitle
%
%________________________________________________________________

\section{Introduction}
\label{s1}

   Kepler satellite marks a revolution in the field of extra-solar
planet study \citep{borucki10}. Apart from thousands of `normal' transiting
exoplanet candidates showing periodic, non-variable, and symmetric
dips, a small number of exceptional transit-like signals were detected. Using data
from this satellite, \cite{rappaport12} discovered a transiting
disintegrating exoplanet KIC12557548b (KIC1255). Unlike all other
exoplanets, it exhibits a strong variability in the transit depth. On
average, transits are approximately 0.6\% deep but they may exceed 1\% or
disappear for some period of time. The shape of the transit is highly
asymmetric with a significant brightening immediately before the eclipse with a
sharp ingress followed by a smooth egress. The planet also has an
extremely short orbital period of approximately 16 hours.

   The interpretation of the light-curve is that the planet is in
a stage of catastrophic evaporation \citep{perez13}. This creates a
comet-like dusty tail extending well beyond the planet's Hill radius
that is responsible for the observed variable transits. The planet
itself is too small to be seen in transit. The mass of the planet must
be relatively small too, less than that of Mars, otherwise the material
would not be able to escape from its deep gravitational well. This is
supported by \cite{garai14} who found no evidence for the dusty tails in
other more massive close-in exoplanets observed by Kepler. The planet's
tail is dominated by radiative and gravitational forces as well as an
interplay between the grain condensation and evaporation. There are also
indications that stellar activity may affect the behavior of the dusty
tail since a quasi-periodic long term variability in the tail
\citep{budaj13} as well as a correlation of the transit depth with the
rotation period of the star \citep{kawahara13} were detected.
\cite{Croll15} argue that the modulation of the transit with the
rotation period may also occur as a result of the star spot
occultations. Pre-transit brightening as well as the color dependence of
the transit depth can constrain the particle size of dust grains in the
comet-like tail, which was found to be of the order of 1 micron
\citep{brogi12, budaj13, croll14, werkhoven14, bochinski15,
schlawin16}.

   Two other objects of this kind (KOI-2700b, K2-22b) were discovered
already by \cite{rappaport14} and \cite{sanchis15}. Their orbital
periods are less than one day. The average transit depth of KOI-2700b is
approximately 0.04\% and decreases with time. Transits of K2-22b
(EPIC201637175B) are 0$-$1.3\% deep. The observed tail lengths in these
objects are consistent with corundum dust grains \citep{lieshout14,
lieshout16}. It has been argued and demonstrated that dust clouds
associated with such exoplanets may not be uniform and may consist of
several structures that may differ in dust properties and chemical 
composition. For example, KIC1255 may have an `inner tail' (coma) and an
`extended trailing tail' (tail) \citep{budaj13,werkhoven14}. A leading
tail may also occur, for example, in K2-22b, if the star is cool and the
radiative acceleration on dust is negligible compared to gravity
\citep{sanchis15}.

   Another exotic object with similar dips in the light-curve is a white
dwarf, WD 1145+017 \citep{vanderburg15, croll15b, xu15}. It features
semi-periodic eclipses with periods of approximately 4.5 hours, which are as deep
as 40\%. A probable explanation is that these dips are caused by
disintegrating planetesimals or asteroids transiting the star. They
gradually fall onto the white dwarf contaminating its atmosphere with
heavy elements. The transits are highly variable, with timescales of
days. The light-curve contains many sharp features that drift if
phased with the dominant period \citep{gansicke16,rappaport16}. 
\cite{zhou16} carried out simultaneous optical and near infrared
observations and placed the lower limit of 0.8 micron on the particle
size of dust grains.

   Recently, an even more peculiar object was discovered in the Kepler
data by \cite{boyajian16} named KIC8462852 (KIC8462). KIC8462 is a 12
magnitude main sequence F3V/IV star with mass of approximately 
$M_{\star} = 1.43\,$M$_{\odot}$, radius $R_{\star} = 1.58\,$R$_{\odot}$,
effective temperature $T_{\rm eff}=6750\,$K, and luminosity 
$L_{\star} = 4.7\,$L$_{\odot}$.  It exhibits the asymmetric variable 
dips in the brightness similar to the above mentioned objects. However,
here the dips have irregular shapes with a tendency towards a smooth
long ingress and sharp egress. Moreover, the dips are much deeper,
sometimes eating up more than 20\% of the flux, and do not show any
obvious periodicity. There is no simple explanation of such behavior.
A first view with GAIA also indicates that it is a normal F3V star
at a distance of $390\,$pc \citep{hippke16c}.

   Based on the analogy with the above mentioned objects, and the strong
extinction properties of dust, one can assume that eclipses are due to
large and opaque dust clouds passing in front of the star. Such dust
clouds may be associated with various objects and/or events. 
\cite{boyajian16} have considered several scenarios; (1) a collision within 
an asteroid belt or planet impact; (2) dust enshrouded planetesimals; 
and (3) the passage of a family of exocomet fragments, all of which are 
associated with a single previous breakup event. The latter scenario seems 
to be the most  consistent with the data. However, it falls short of explaining the shape of the dips.

   Infrared observations using WISE, Spitzer, and NASA/IRTF 3m
\citep{boyajian16,marengo15,lisse15} have not detected any significant
infrared excess emission, which puts strong constraints on the presence
and amount of hot dust in the vicinity of the star. Aside from this non-detection, no
significant emission from cold dust was detected at millimeter and
sub-millimeter wavelengths by \cite{thompson16}, which limits the amount
of dust within $200\,$au from the star to less than $7.7\,$M$_{\oplus}$
and amount of dust actually occulting the star to less than approximately
$10^{-3}\,$M$_{\oplus}$.

   \cite{bodman16} investigated the possible comet scenario and found
that it is possible to fit most of the features in the Kepler
light-curve with several clusters of comets containing 70$-$700 comets.
However, it was not possible to reproduce a large dip at day 800
due to its smooth shape and gradual ingress followed by a sharp egress.

   It was pointed out by \cite{wright16} that
% \LEt{I would suggest that the numbering of various scenarios should
% be stopped at three and continuous prose be resumed. This is in line
% with AandA guidelines concerning numbering. AandA prefers that authors
% avoid bullet points and numbering. Please take out all following
% numbers from the text in this section }(4)
the above mentioned kind of variability might be consistent with a
`swarm' of artificial mega-structures produced by an extraterrestrial
civilization. \cite{harp16} and \cite{schuetz16} searched for the
presence of radio and optical signals from extraterrestrial intelligence
in the direction of the star and found no narrow band or wide band radio
signals or periodic optical signal. \cite{abeysekara16} also searched
for brief optical flashes towards the target and found no evidence of
pulsed optical beacons above a pulse intensity at the Earth of
approximately 1 photon per m$^2$.

   The recent discovery of a long-term fading of the star by 0.16 mag
over the last century makes the situation even more complicated and 
also poses a problem for the comet scenario \citep{schaefer16}. 
`Fortunately', as pointed out by \cite{hippke16}, \cite{lund16}, and
\cite{hippke16b}, this long-term trend is most likely a data artifact,
and it is probably not of astrophysical origin. Nevertheless,
\cite{montet16} found convincing evidence from the Kepler data that the
star had dimmed by approximately 3\% during the duration of the Kepler mission.
In connection with this, it was suggested that % (5)
the variability may be due to free-travelling interstellar material in
the form of either a dark cloud (Bok globule), a disk around a stellar
remnant, a swarm of comet-like objects, or planetesimals crossing the
line of sight \citep{wright16b,makarov16,lacki16}. \cite{lisse_etal16}
suggest % (6)
a parallel between KIC8463 and the observations of a late, heavy
bombardment of $\eta$ Corvi.

   The aim of the present study is to explore the shapes of the eclipse
events observed in KIC8462 and investigate whether or not it is possible
to comprehend their basic features in terms of eclipses of only a small
number of
% \LEt{the term `a few' is overly vague here.}
massive bodies and associated dust clouds (hence a small number
of free parameters). We do not aim to determine the parameters of the
dust and parent bodies unambiguously but rather provide hints for future
more detail investigations.

\section{Observations, motivation, and aims}
\label{s2}

\begin{figure*}
\centerline{
%\includegraphics[width=8.2cm, angle=-0]{event0800.pdf}
%\includegraphics[width=8.2cm, angle=-0]{event1520.pdf}}
%\centerline{
%\includegraphics[width=8.2cm, angle=-0]{event1540.pdf}
%\includegraphics[width=8.2cm, angle=-0]{event1570.pdf}}
\includegraphics[width=5.2cm, angle=-90]{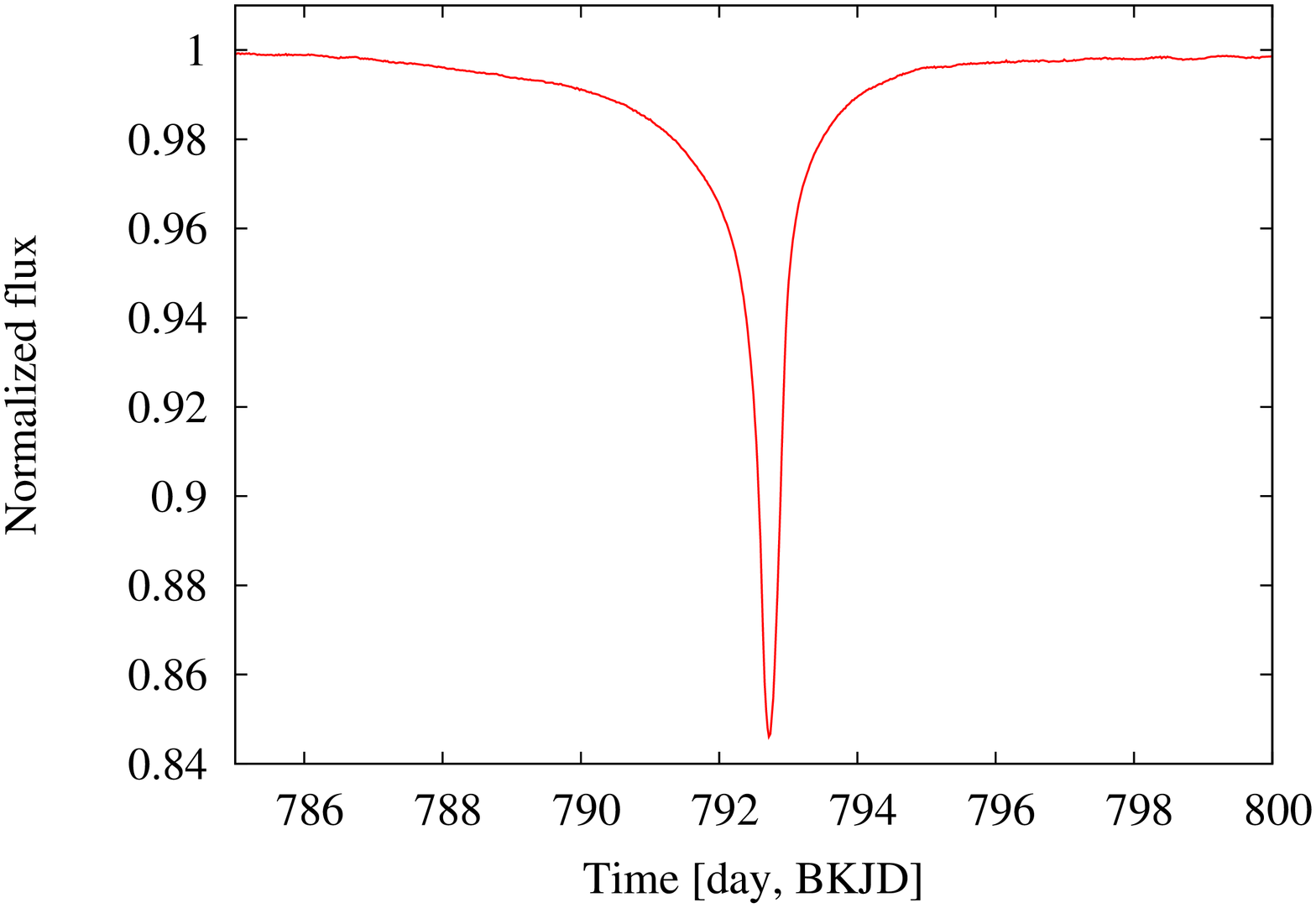}
\includegraphics[width=5.2cm, angle=-90]{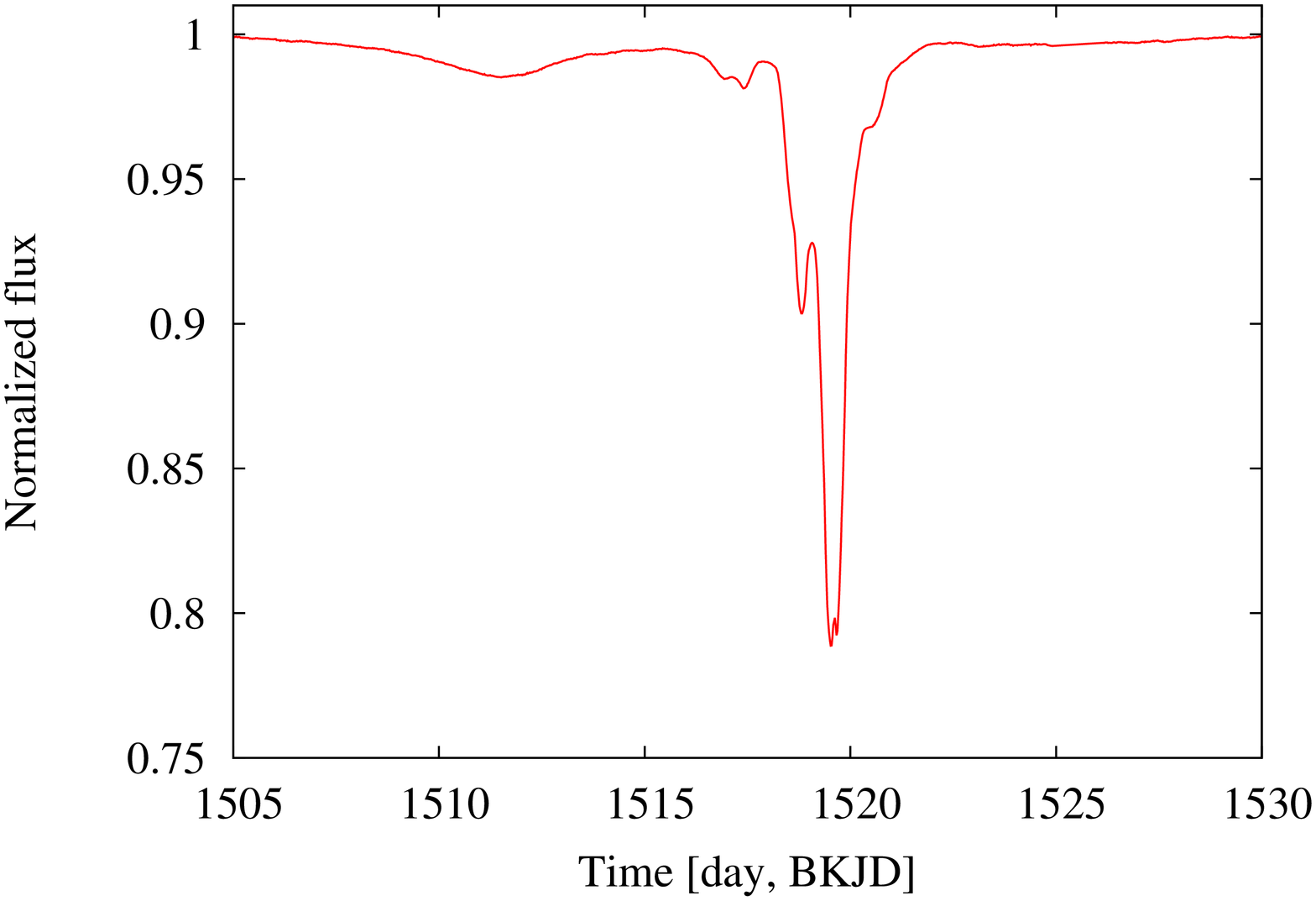}}
\centerline{
\includegraphics[width=5.2cm, angle=-90]{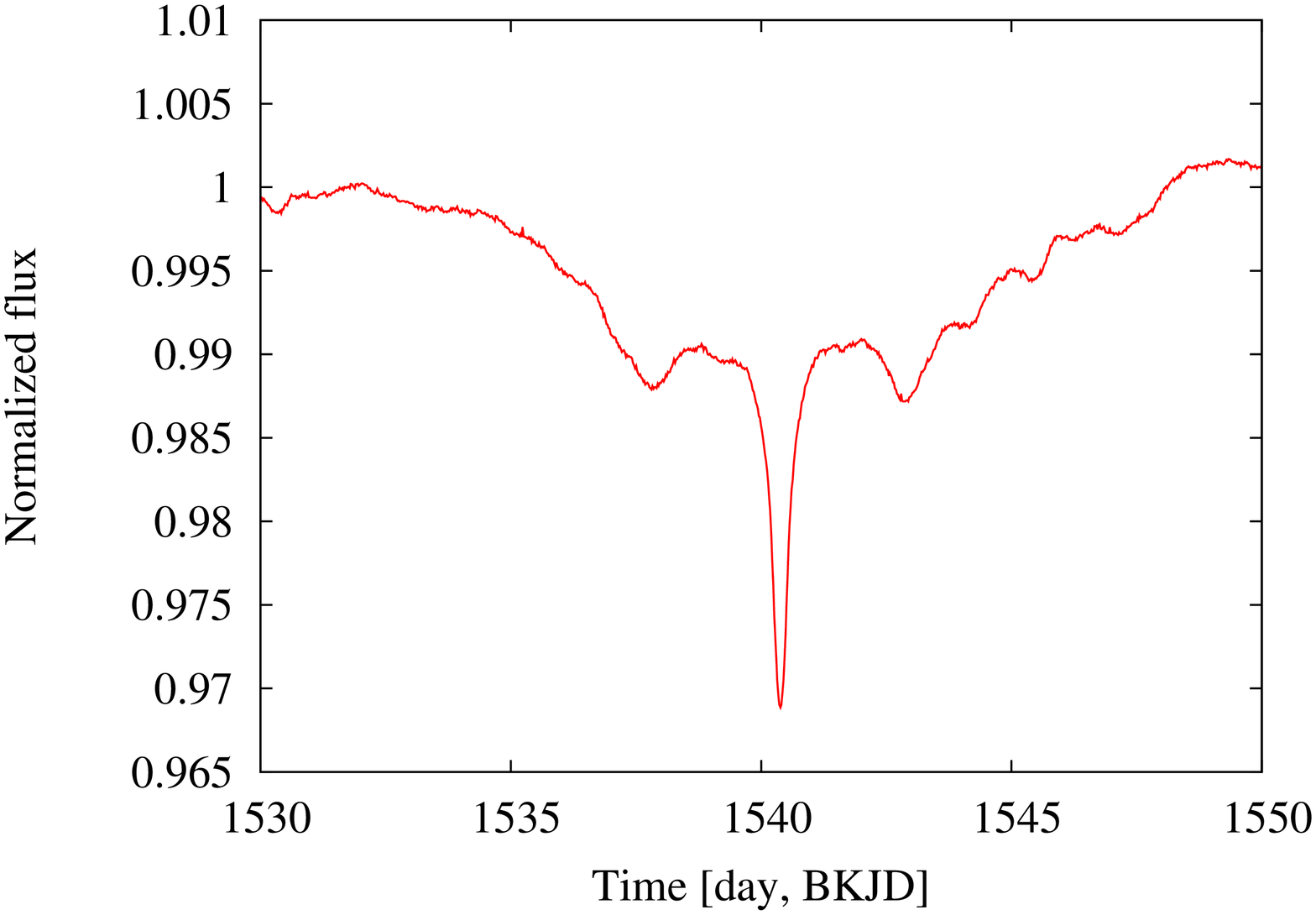}
\includegraphics[width=5.2cm, angle=-90]{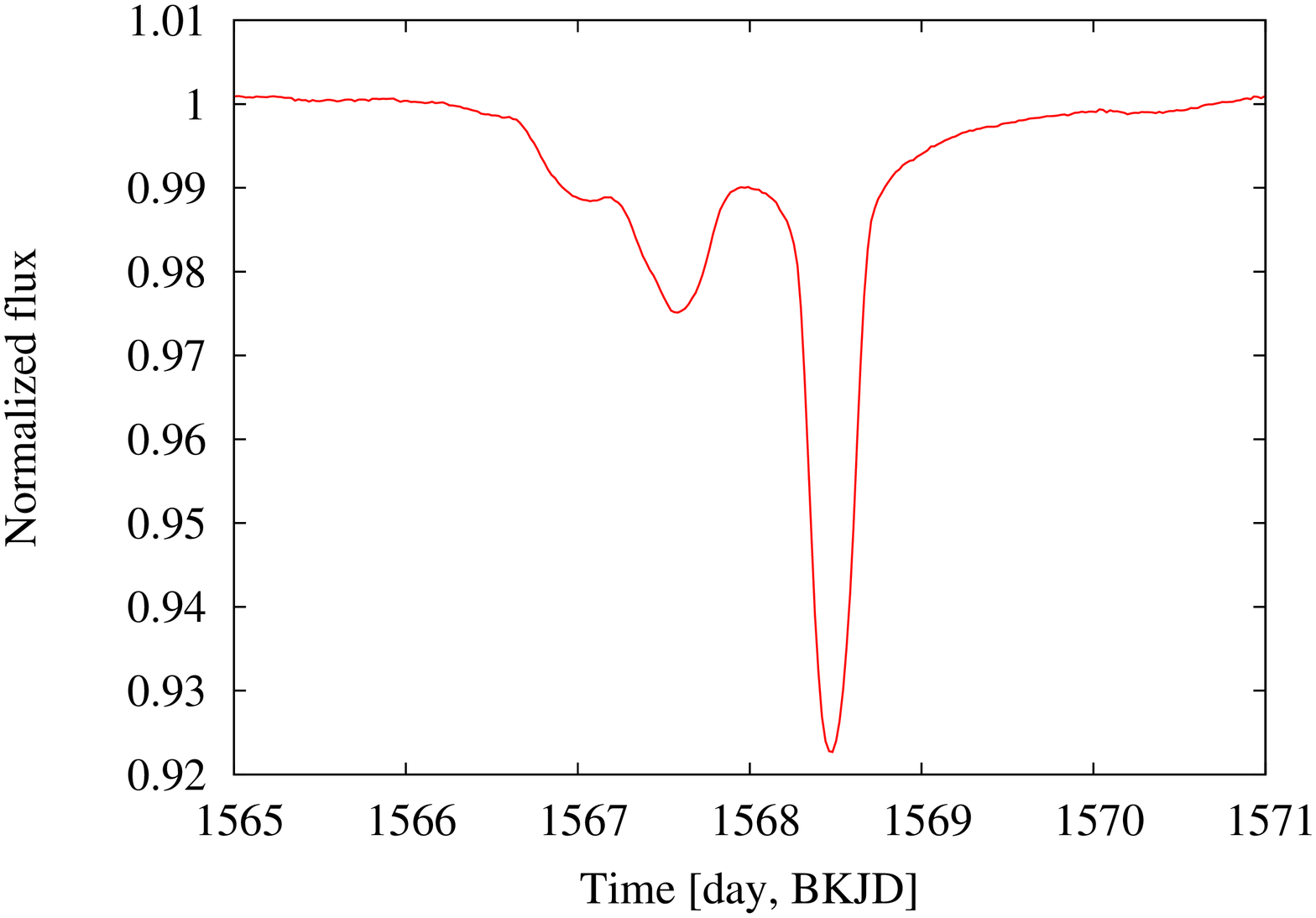}}
\caption{Four main eclipse events in the light-curve of KIC8462 observed
with KEPLER \citep{boyajian16} and also analyzed in this study.}
\label{f1}
\end{figure*}

   We use the data for KIC8462852 obtained by the Kepler satellite 
\citep{borucki10}. These are 18 quarters of long cadence observations in
the form of PDCSAP flux as a function of Kepler Barycentric Julian Day
(BKJD). The exposure time of long-cadence observations was approximately 30
minutes. There are significant offsets between the data from different
quarters, and for this reason, each quarter was normalized separately.
Normalization was very simple; PDCSAP flux from each quarter was divided
by a single constant. The advantage of such a simple normalization is
that it does not introduce any artificial trend to the data. 
Four main events can be observed in this light-curve around
BKJD of 800, $1\,520$, $1\,540$, and $1\,570$, respectively.
They are shown in Fig.~\ref{f1}. 
Aside from that there are other relatively small transit events for 
example at 140 and 260 days with a depth of approximately 0.6\%.
% \LEt{perhaps this would be obvious to those in your field, but you may
% want to add units here.}.
These are not the subject of this study
but this information is used later in the discussion.

% FIG. 1

   As already mentioned above, even the most promising comet scenario
\citep{bodman16} fails to explain all of the recorded observations. Most of the observed dips (see
Fig.~\ref{f1}) tend to have a slow ingress followed by a sharp egress.
This is in contrast with a typical transit of a comet-like body, which
would show a steep ingress followed by a slow egress \citep{lecavelier99}. The observed light-curves, if explained by comets, would
require comets in large numbers, larger than $10^2$, and consequently
many free parameters. Furthermore, it is
difficult to reproduce, for example, the egress part of light-curves. Some
features cannot be represented at all, such as a smooth feature at 800
days. Some others, such as a symmetric triple feature at 1$\,$540 days,
might suggest the presence of a ring-like structure that would imply an
object with a non-negligible gravity. This motivates our search for an
alternative explanation of the phenomenon. Our goal is to investigate
whether or not it is possible to reproduce the observed features with only a
small number of eclipsing objects, based on first principles and simple assumptions.
Of course one cannot expect a perfect fit with only a handful of free
parameters.

   The transit of a dark spherical opaque body can cause a relative drop
in the light-curve as deep as $R_{p}^{2}/R_{\star}^{2}$ where $R_{p}$ and
$R_{\star}$ are radii of the body and star, respectively. No single
planet or brown dwarf could cause an asymmetric 20\% dip in the
light-curve of a solar type main-sequence star. That is why we will
assume a massive object (MO) surrounded by a dust cloud. It would be this
dust cloud that would cause the eclipse events. We refer to a `massive object'
as an object with a non-negligible gravity (heavier than a
large comet in the Solar System) so that it has to be taken into
account.

   Here, `dust cloud' refers to a cloud containing numerous individual massless
dust particles (DPs), which initially reside within the gravity well of
the parent body. This helps to keep the DPs together to form a cloud
associated with the parent body. One could argue that (A) the
probability of a transit rapidly decreases with the distance of the
cloud from the star indicating that the star-cloud distance during the
transit may be rather small. On the other hand, (B) the dust cloud will
radiate in the optical region (scattered light) as well as in the
infrared and mm region (thermal radiation). This radiation diminishes with approximately the square of the star-cloud distance, and, as mentioned in
the Sect.\ref{s1}, observations of the thermal radiation exclude large 
amounts of the dust located close to the star at the time of
observations. It is possible to adhere to both constraints (A) and (B) if
a dust cloud is located on a highly eccentric orbit. The same idea and
arguments are used in the comet scenario and are therefore also acceptable in
this scenario.

   Ultimately, a few massive objects with dust clouds would be needed to
describe the few complex features seen in the light-curve. 
The orbits of these objects should be situated in the same
plane  and have similar trajectories. Notice that although we go on to
assume approximately four massive objects on almost identical orbits, these
objects are not `statistically' independent. They are apparently of a
common origin, that is, the result of a break-up process. Thus, the statistical
probability of our model is reduced to the existence of a single massive
body on an eccentric orbit and a break-up process.

   The origin of the dust clouds and their parent bodies is not the
subject of this study and their presence and initial structure are both assumed for the purpose of this study. Nevertheless, one can speculate that they might be the result
of an impact, collision, or break-up event facilitated by the rotation,
tides, or heating from the star. For example, a recent study of
catastrophic disruption of asteroids in the Solar system by
\cite{granvik16} shows that asteroids break up at much larger distances
from the Sun than previously thought; approximately $0.094\,$au. If this
value were scaled by $\sqrt{L_{\star}/L_{\odot}} \approx 2.2$, asteroids
around KIC8462 would break up at a distance of $0.20\,$au, which is
considerably larger than its Roche limit, which is approximately
$2.44\,$R$_{\star}$ or $0.018\,$au (assuming that the density of the
star and that of the asteroid are approximately the same). The low-albedo asteroids
break up more easily and at larger distances than high-albedo asteroids.
Another possibility is that rocks break into small grains by thermal cracking,
or the YORP effect causes the asteroids to spin faster, to the point
when gravity and cohesive forces can no longer keep them intact
\citep{vokrouhlicky15}. A third possibility is that all asteroids
contain volatile elements that, when sublimating at moderate
temperatures, exert enough pressure on the body to cause it to explode.

   Discovery of KIC8462 in the Kepler data may be used to estimate
the frequency $f$ of such systems under the assumption that they are
associated with a massive body on a highly eccentric orbit,  such as
the one used in our calculations, for example, with a periastron of approximately $0.1\,$au
and an apastron of approximately $50\,$au. The number of such events, $N_{obs}$,
observed during the duration of the Kepler mission is:
 \begin{equation}
N_{obs} \approx N_{star} f P \frac{T}{P_{orb}}
 ,\end{equation}
where $N_{obs}=1$ is the number of such events
discovered with Kepler. $N_{star}$ is the number of stars monitored with
Kepler (approximately $10^{5}$) (main sequence stars brighter
than approximately $m_{V}=14$ mag.). $P$ is the probability that the body has a
proper inclination and argument of periastron to detect the transit,
which is approximately 0.01$-$0.1 for the assumed body. $T/P_{orb}$ is the
probability of catching the transit. $T=4\,$yr is the duration of the
Kepler mission and $P_{orb}\approx10^{2}\,$yr is the orbital period of
the assumed body. From the above equation, the probability that a star
currently hosts pieces of such a broken body with dust clouds that are
capable of producing such events is $f\approx 10^{-2}$.

   The existence of a massive body on a highly eccentric orbit may
not be highly unusual. In our own Solar System, some comets
from the Oort cloud \citep[e.g.,][]{duncan87,neslusan05} as well as
trans-Neptunian objects \citep{emelyanenko07} have been found to migrate
down to the region of terrestrial planets. Their perihelion can be
reduced to a few astronomical units or even closer. The aphelion of such
an object can remain beyond the orbit of the outermost planet. According
to \cite{bailey92}, if the orbital inclination of the objects entering
the inner Solar System is high, the perihelion distance of many of them
is further reduced due to the long-term secular resonances and the
objects become sun-grazing at intervals of approximately $10^{3}$
revolutions. The authors introduced five representative comets currently
evolving to become the sun-grazers: 96P/Machholz, 161P/Hartley-IRAS,
C/1846 B1, C/1989 A3, and C/1932 G1.

   Close to the periastron located in close vicinity to the
central star, the object can be destroyed by the tidal action of the star.
A remnant of such destruction in the Solar System can still be detected
in the form of small sun-grazing comets orbiting the Sun along similar
orbits. Specifically, four such groups of sun-grazing comets
are observed in our Solar System at present; Kreutz, Meyer, Marsden, and Kracht group \citep{biesecker02,knight10}. Each group is believed to
originate from a single, more massive progenitor. The most highly populated is
the Kreutz group. Besides the small cometary nuclei, bright comets have also been observed within this group, providing evidence that
some large fragments still exist in it \citep[e.g.,][]{sekanina15}.
Consequently, the occurrence of a massive body on such an eccentric
orbit at any time is not rare. If there are groups of comets, each
group moving in similar orbits and originating from a single progenitor
in our own Solar System, there could also be a group of objects in almost
identical orbits at the KIC8462.

   Perhaps it is worth noting that the similarity of the orbits
of group members (orbits of the MOs) occurs due to the small velocity of
the separation of the members from their progenitor at its break-up in
comparison to the orbital velocity of the progenitor itself.

\section{Dust properties}
\label{s3}

\begin{figure*}
\centering
\includegraphics[angle=-90,width=8.5cm]{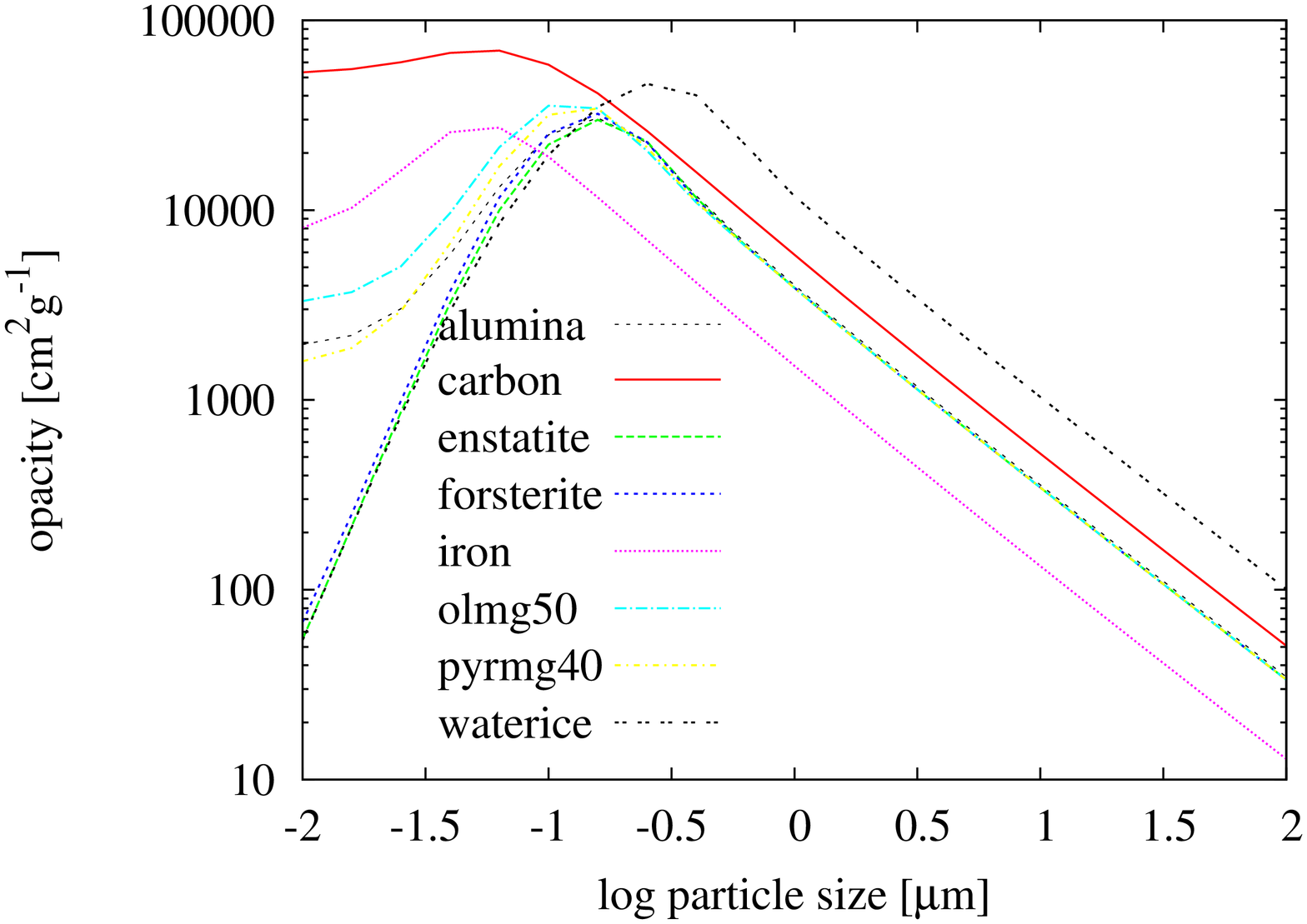}
\includegraphics[angle=-90,width=8.5cm]{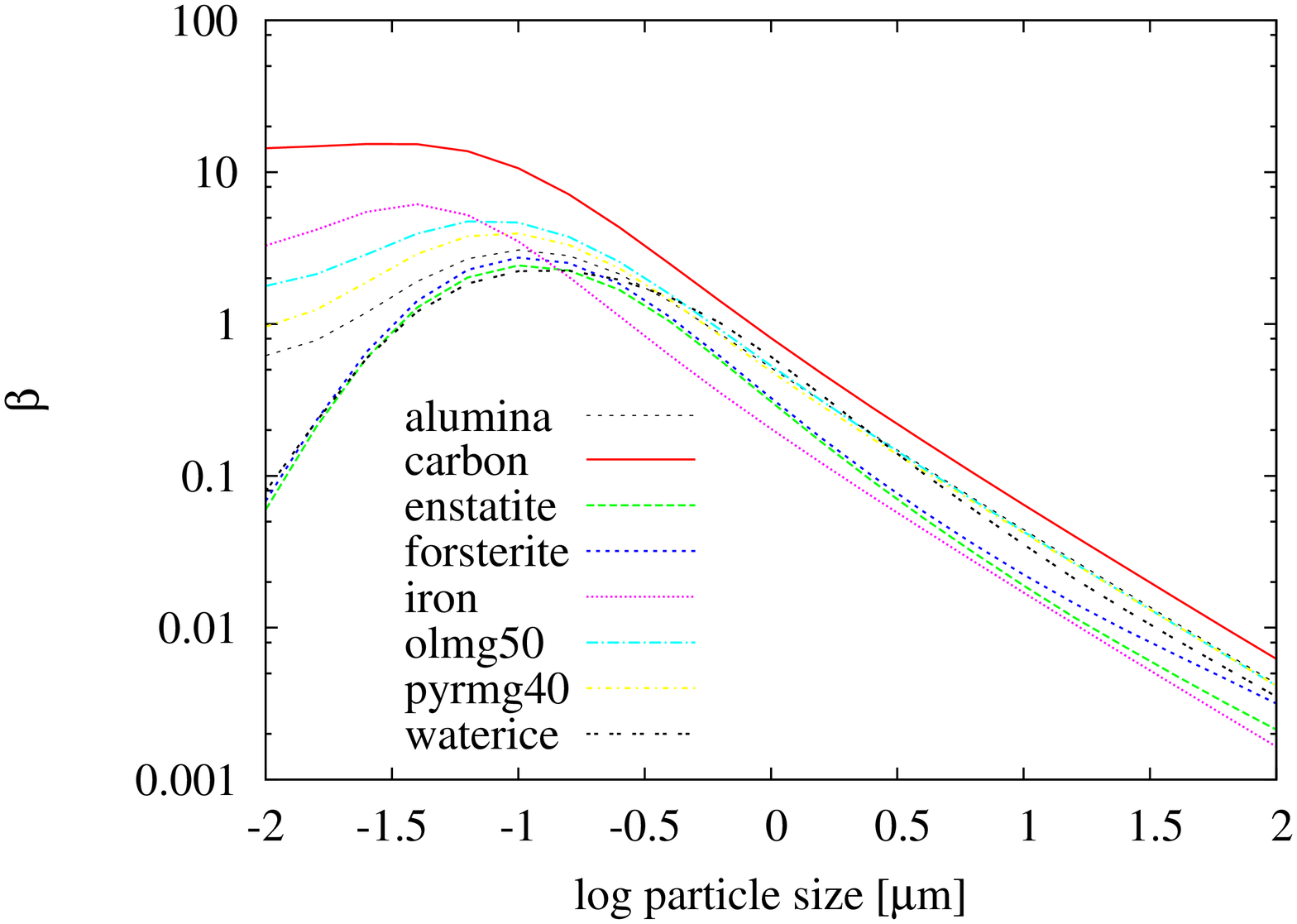}
\caption{Left: total opacity of various dust species at 0.6 microns as
a function of particle size. Right: radiative to gravity acceleration
ratio {\bf($\beta$ parameter) }of various dust species for KIC8462 as
a function of the particle size. Olmg50 refers to olivine  (50\%
iron), pyrmg40 refers to pyroxene (60\% iron), and carbon is at
$1000\,^{o}$C.}
\label{opac}
%\label{beta}
\end{figure*}

   The shape and depth of obscuration events will depend on certain
properties of dust grains. It is mainly the number of dust particles
along the line of sight and their size that determines the opacity and
optical depth along the line of sight. This is illustrated in
Fig.~\ref{opac} for a number of refractory dust species and for a wavelength
of approximately 0.6 microns corresponding to the Kepler spectral window. The
opacities include both absorption and scattering opacity and are in
cm$^{2}\,$g$^{-1}$ , which means that they are per gram of the dust
material. One can see that the opacity of most species has a pronounced
maximum at approximately 0.1 microns.

   As shown, it takes much more dust at low opacity to produce
the same dip in the light-curve compared to dust at high opacity. Table
\ref{t1} lists the minimum amount of dust required to produce a 20\% deep
eclipse of KIC8462 at 0.6 microns as a function of the particle size for
iron, forsterite (an iron free silicate of the olivine family), and
liquid water, assuming an optically thin dusty environment. No other
assumption is involved in this estimate. The amount of dust required in
the form of even larger particles is linearly proportional to the
size of particles and can be easily extrapolated.

% TAB. 1
\begin{table}
\caption{Minimum dust mass in M$_{\oplus}$ for different dust
compositions. 
Particle size is the radius of a grain in microns.}
\centering
\begin{tabular}{c c c c}
\hline\hline               
Size & Iron          & Forsterite    & Water        \\ 
\hline                      
0.01 & 2$\times10^{-10}$ & 2$\times10^{-8} $ & 2$\times10^{-8} $\\ 
0.1  & 7$\times10^{-11}$ & 5$\times10^{-11}$ & 6$\times10^{-11}$\\
1    & 8$\times10^{-10}$ & 3$\times10^{-10}$ & 1$\times10^{-10}$\\
10   & 1$\times10^{-8}$  & 4$\times10^{-9} $ & 1$\times10^{-9} $\\
100  & 1$\times10^{-7}$  & 4$\times10^{-8} $ & 1$\times10^{-8} $\\ 
\hline      
\end{tabular}
\label{t1}
\end{table}

   Another important property of dust grains is that apart from gravity
of the MO and the star, they may experience significant acceleration due
to the radiation of the host star. This radiative acceleration is also
controlled by the absorption and scattering opacities of the grain,
angular distribution of the light scattered by the grain, and spectral
energy distribution, size, and distance of the star. This is also
illustrated in Fig.~\ref{opac} as a function of particle size
in the form of  the $\beta$ parameter, which is the radiative to gravity
acceleration ratio. Since the mass of KIC8462 and, hence, its gravity is
larger than that of the Sun, one must remember that the same value of
$\beta$ means a larger radiation pressure at KIC8462 than at the Sun.

   The $\beta$ parameter also shows a maximum at approximately 0.1 microns,
which reflects the peak in opacity. Because the radiative acceleration
upon a grain is proportional to its cross-section while gravity is proportional to its
volume, very large and heavy grains have $\beta \ll 1$. In such a case, the
radiative pressure can be neglected and, in the absence of other massive
objects, grains remain on an elliptical orbit around the star. For
grains with $\beta=1$, gravity from the star will be balanced by the
radiation pressure and, once the grain escapes the gravity of the
planetesimal, it follows an almost straight line. Grains with
$\beta>1$ and `out of reach' of the MO will follow a hyperbolic orbit
around the star `forbidding them from passing close to the star or revolving around it; such grains are expelled from the system.
Consequently, the chance of causing an occultation event is significantly
lower for these grains unless they are constantly replenished.
This is the case for dust grains with a size of approximately 0.1 microns
as well as smaller carbon, iron, or iron-rich grains. This is shown in
Fig.~\ref{betaorb} assuming a model A of the cloud (see
Sect.~\ref{calculations} below) around the MO with the mass
$m=10^{-10}\,$M$_{\star}$. The plots (a), (b), (c), and (d) illustrate
 the motion of the DPs calculated for $\beta=0.007$, 0.7, 1, and 1.4,
respectively.

% FIG. 2

% FIG. 3
\begin{figure*}
\centerline{\includegraphics[angle=-90,width=8.5cm]{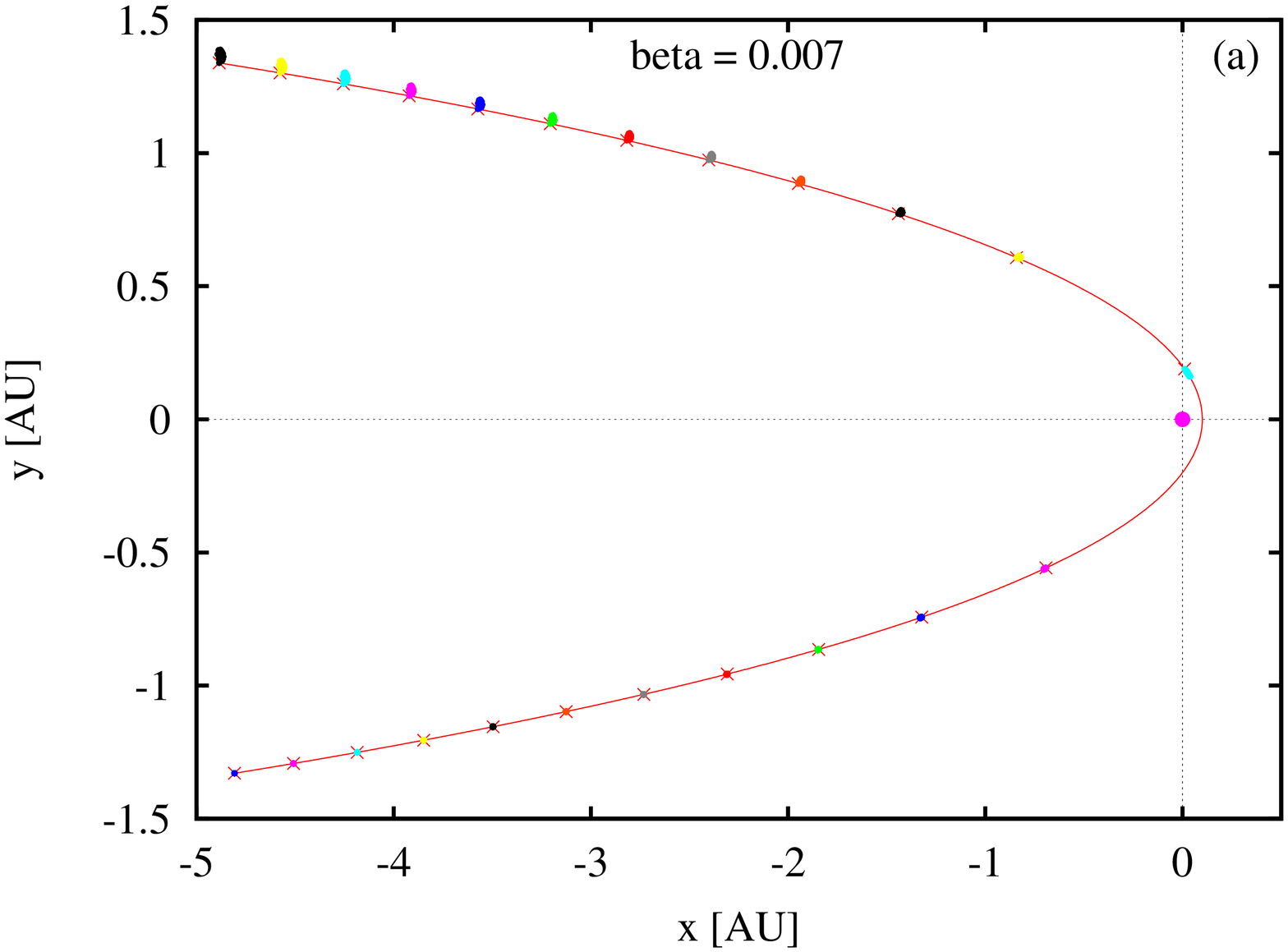}
            \includegraphics[angle=-90,width=8.5cm]{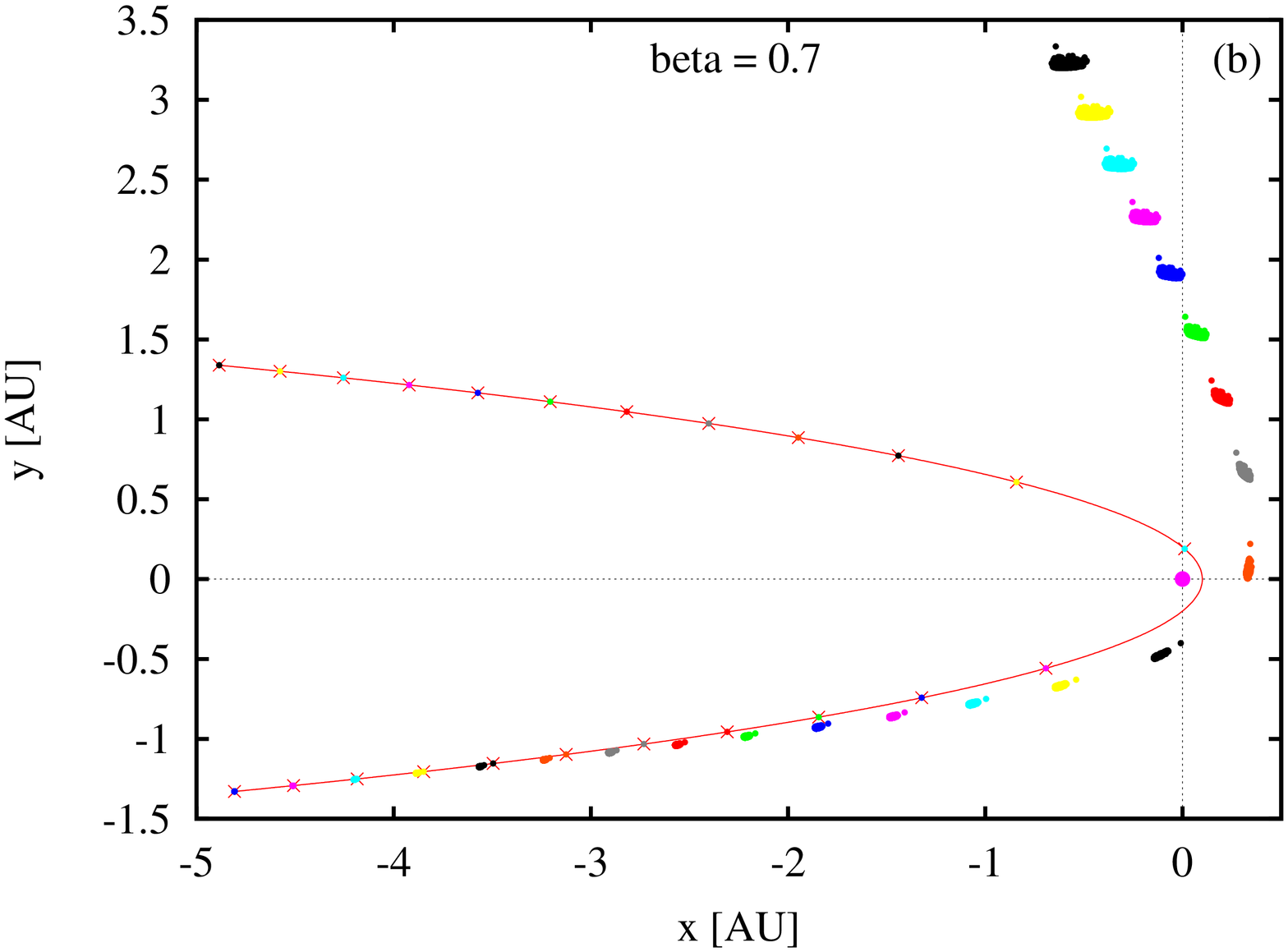}}
\centerline{\includegraphics[angle=-90,width=8.5cm]{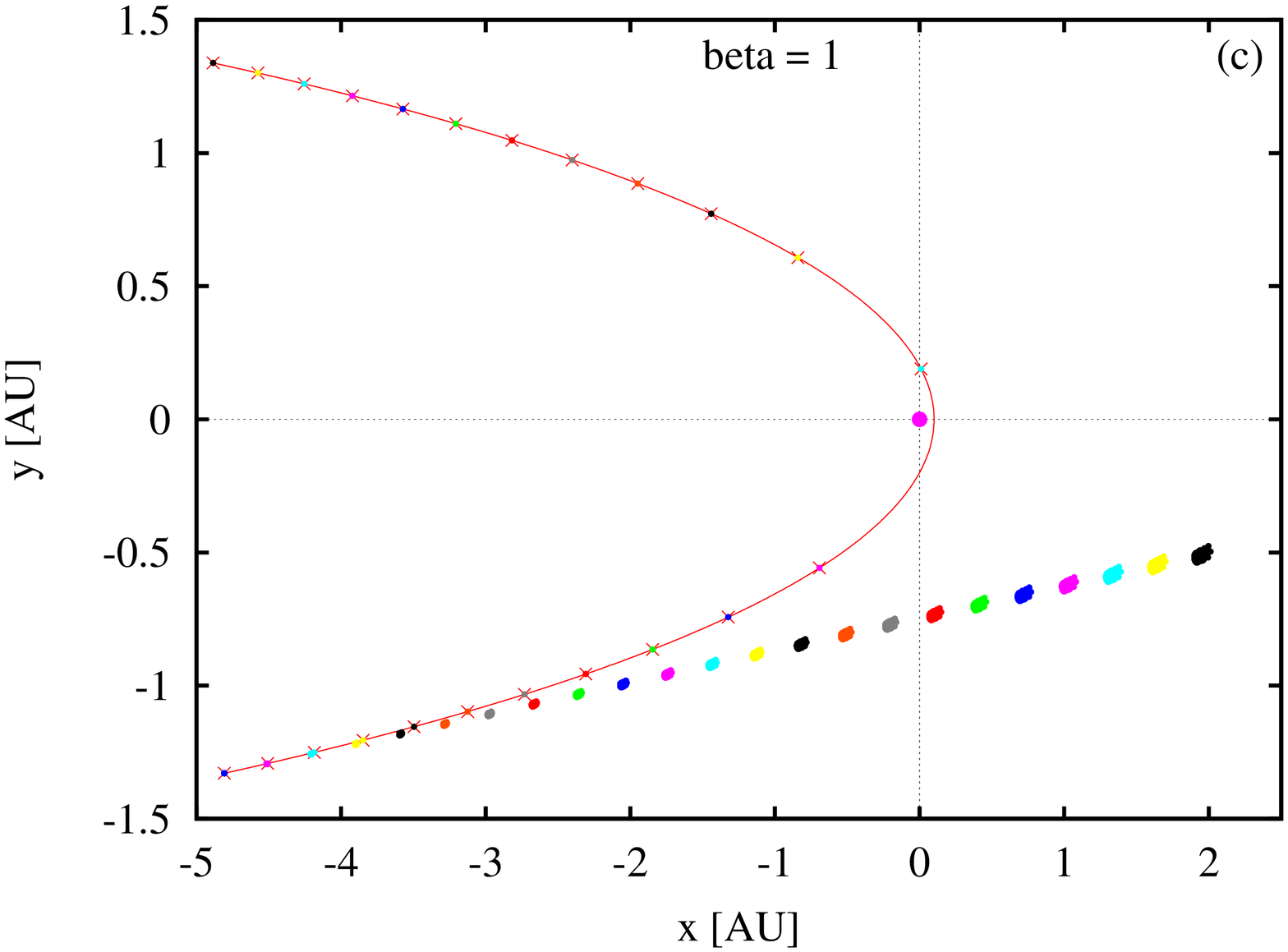}
            \includegraphics[angle=-90,width=8.5cm]{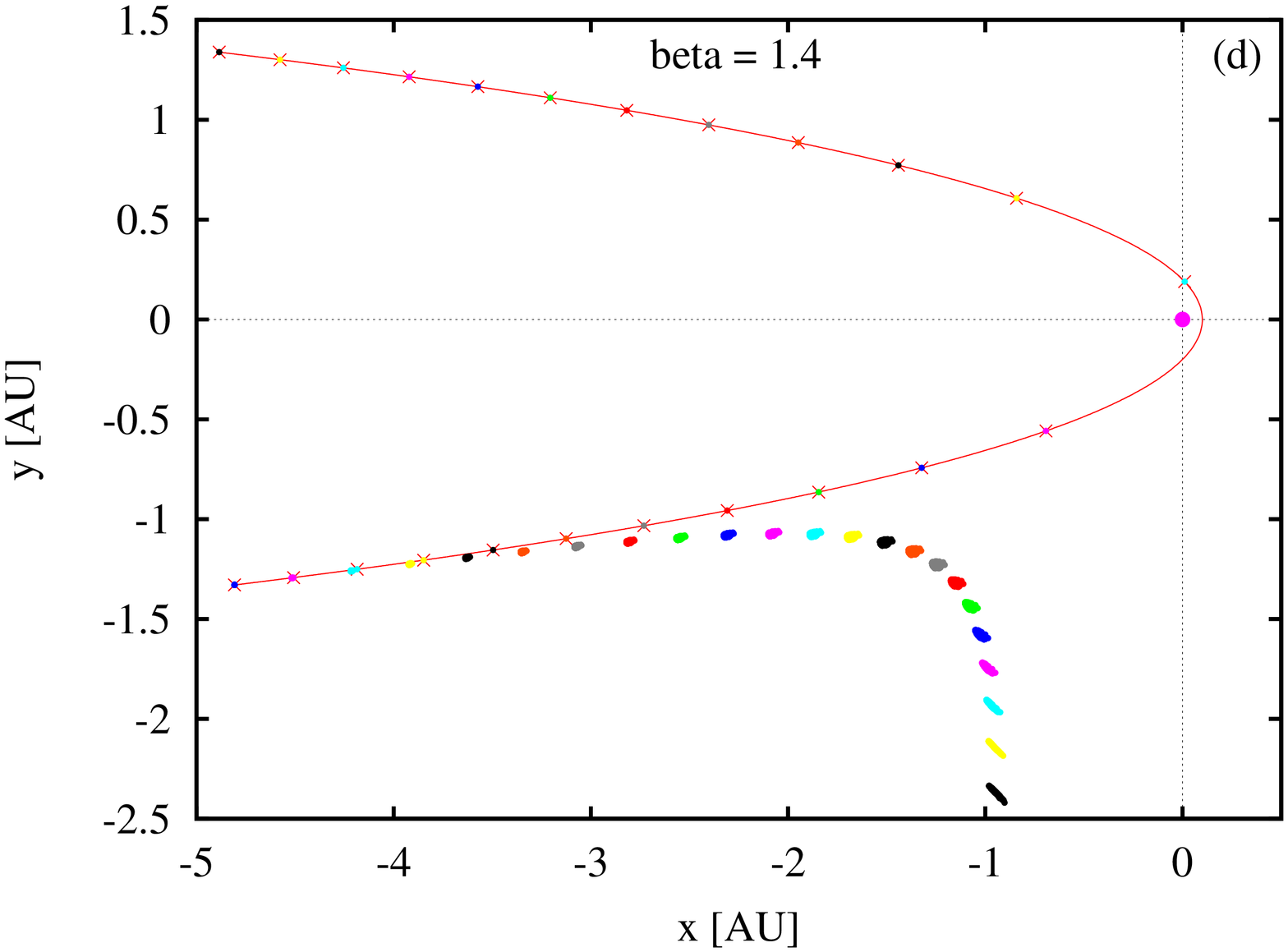}}
\caption{Illustration of the behavior of a dust cloud projected into the
orbital plane of the MO for different $\beta$ values. Plot (a):
$\beta = 0.007$ - particles tend to stay on the elliptical orbit and
follow the parent body, the trajectory of which is shown with the red
solid curve. Plot (b): $\beta = 0.7$ - the cloud of particles soon
separates from the parent body and the particles move in their own elliptical
trajectories. Plot (c): $\beta = 1$ - the P-R drag balances the gravity and
therefore the particles move uniformly along a straight line. Plot (d):
$\beta = 1.4$ - particles easily decouple from the parent body and settle
onto hyperbolic orbits. Positions of the MO (crosses) are shown at the
same moments as the positions of the individual DPs (dots). Different
colors are used for different times, separated by 25-day intervals. The
position of KIC8462, in the origin of coordinate $x$$-$$y$ plane, is
drawn with the violet full circle. Calculated for model A of a cloud with
a body mass of $m=10^{-10}\,$M$_{\star}$.}
\label{betaorb}
\end{figure*}

   Given the above mentioned accounts, opaque grains with $\beta<1$
values are the best candidates for causing such obscuration events. This is
fulfilled for grains with $0.1<\beta<1,$ which translates to
particle sizes of approximately 0.3$-$10 microns for most of the dust species.
Based on Table~\ref{t1} it would require at least $10^{-10}$ Earth masses of
dust. The above mentioned calculations made use of the on-line tables of
dust properties calculated by \cite{budaj15}. They assume homogeneous
spherical grains with a relatively narrow Deirmendjian particle-size
distribution. Particle size of such a distribution refers to its modal
particle size. Radiative accelerations assume non-black body radiation
from the star\footnote{BT-Settl models \citep{allard03,baraffe15}} with effective temperature, mass, and radius of $T_{\rm eff}=6750\,$K,
$M=1.43\,$M$_{\odot}$, and $R=1.58\,$R$_{\odot}$, respectively
\citep{boyajian16}. Table~\ref{t2} lists the sources of the refractive
index used in the calculations.

% TAB. 3
\begin{table}
\caption{Reference list for the refraction index used in this work.
Olivine50 refers to olivine with 50\% iron content, pyroxene40
refers to pyroxene with 60\% iron, and carbon1000 refers to carbon with a
temperature of $1000\,^{o}$C.}
\label{t2} 
\begin{tabular}{ll}
\hline
species    & reference   \\
\hline
alumina    &  \cite{koike95}                   \\
iron       &  \cite{johnson74},\cite{ordal88}  \\
forsterite &  \cite{jager03}                   \\
olivine50  &  \cite{dorschner95}               \\
enstatite  &  \cite{dorschner95}               \\
pyroxene40 &  \cite{dorschner95}               \\
carbon1000 &  \cite{jager98}                   \\
water ice  &  \cite{warren08}                  \\
\hline
\end{tabular}
\end{table}

\section{Calculations of the obscuration events}
\label{calculations}
\subsection{Model of the dust cloud}

   Light-curves of the obscuration events depend on the properties,
position(s), and velocity(ies) of the parent massive body and the individual dust
grains, as well as the position of the observer. These are all unknown
quantities. Furthermore, there is an infinite number of possible models of a
dust cloud to be envisaged and it is impossible to study all of
them in detail. That is why, in this primary study, we restrict our efforts
to a few simple models that can be described by a relatively small number of free
parameters. At the same time we use only a few of the more
important and/or probable values of these free parameters.

   For example, in line with what was argued in the Sect.~\ref{s2}, we
assume a massive object analogous to a destabilized trans-Neptunian
object, orbiting KIC8462 at a highly eccentric orbit. More specifically,
we often consider a `standard orbit' with the apastron equal to
$50\,$au. The periastron is assumed to equal $0.1\,$au, which fits well
with individual features and is still inside the $0.2\,$au destruction
zone mentioned in Sect.~\ref{s2}. For simplicity, an edge-on inclination
of $i=90^{o}$ is assumed. We consider a range of masses from that of a
giant cometary nucleus ($\approx 10^{-6}\,M_{\oplus}$) to approximately four Moon
masses ($\approx 5 \times 10^{-2}\,M_{\oplus}$). Detailed
information about the input parameters in all considered models is
given in Tables~A.1 to A.5 of the Appendix.

% FIG. 4
\begin{figure}
\centering
\includegraphics[width=5cm]{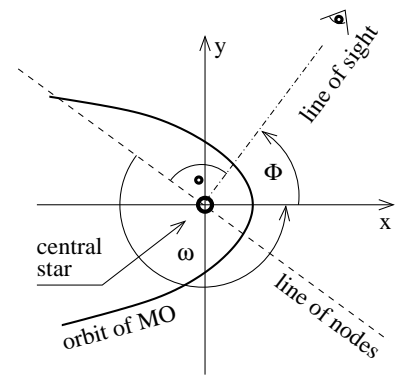}
\caption{Coordinate system in respect to the MO orbit and direction 
towards the observer.}
\label{scheme}
\end{figure}

   Further, we assume a cloud of massless dust particles orbiting the MO.
Their initial locations and velocities are defined at the moment when the
MO is situated at a distance $r_{o}$ from the star, well before the
periastron. The vast majority of models assume $r_{o}=5\,$au. This
choice was motivated by the comet activity in our Solar system due to
sublimation of water ice at a distance of approximately 2$-$3$\,$au from the Sun
\citep{delsemme71}. Scaling this distance by a factor of
$\sqrt{L_{\star}/L_{\odot}} \approx 2.2,$ one could expect an onset of 
similar activity of KIC8462 at a distance of approximately $4$$-$$7\,$au.
In a number of test models (see tables in Appendix), the cloud was also
created in the periastron or immediately before it, when the MO reached
a distance
% \LEt{from the? - I would suggest that the author add this information
% at this precise position in the sentence for ease of understanding.}
critical for its splitting. However, the light-curves from these models
did not match, even approximately, their observed counterparts.
This is because DPs did not have enough time to be dispersed.

   A 3D Cartesian coordinate system $O(xyz)$ is used to describe
the orientation of the MO and DPs, and for the calculations. $x$$-$$y$ is the
MO's orbital plane, $x$ axis points towards the MO periastron, $y$ axis
is oriented in the sense of motion of the MO at the periastron, and the $z$
axis is such that one would see an anti-clockwise orbit from the
positive $z$-values $-$ see the scheme in Fig.~\ref{scheme}. The angle
measured from the $x$-axis to the line of sight is denoted by $\Phi$. 
The following three models for the dust clouds, specified by the
initial conditions of DPs, were considered.

%\subsubsection{Model (A)}
\begin{itemize}
\item 
Model (A):
all DPs are placed on initially eccentric orbits around MO. We assume
several specific values of their (unique) pericenters and initial
pericenter velocities (see Tables~A.1$-$A.3). The pericenter ($q_{tp}$)
and apocenter ($Q_{tp}$) distances of the DPs with respect
to the MO were mostly assumed to equal $10^{3}$ and $10^{5}\,$km,
respectively. The orientation of the orbits is random, as is the line of
apsides.

   All DPs are simultaneously released from their pericenters with
similar velocities, meaning similar apocenters. This might be
interpreted as a spherical shell occurring after a sudden release of
dust, outburst, eruption, or explosion on
% \LEt{on or of? please verify.}
the object. The sudden outburst may be common, especially on smaller
objects, such as comets (a good example is the well-known frequently
outbursting comet 29P/Schwassmann-Wachmann 1) or some asteroids, such
as (596) Scheila \citep{warner06,husarik12} or P/2010 A2 
\citep[][and many others]{jewitt09,birtwhistle10}. However, notice that
since all the particles are initially located at their pericenters
and are released simultaneously on elliptical orbits with similar
pericenters and apocenters, this may trigger an oscillation of the cloud.
The oscillation period equals the orbital period of DPs, which depends
on their semi-major axes and the mass of MO. For example, assuming the
masses of the MO equal $10^{-8}$, $10^{-10}$ , and
$10^{-12}\,$M$_{\star}$, and a DP semi-major axis of 50$\,$000$\,$km, a
given particle would reach its apocenter on Keplerian orbit after 9.5,
95, and 950 days, respectively.

   We were originally skeptical about such a simple type of model but,
to our surprise, we found a number of light-curves which resemble
the observed ones.

\item
Model (B): similar to model (A) except that the particles are not released
simultaneously from the pericenter. They initially have a random
distribution of their mean anomalies. At the same time, they have a
random distribution of their initial velocities (corresponding to the
apocenter distances typically in a range from $10\,000$ to
$100\,000\,$km; some other intervals of the apocenters were
also considered, however, (see Table~A.4). The value of the pericenter distance
is still chosen to be the same for all DPs. The random mean anomaly and
initial velocity places them on different orbits around MO with
different orbital periods. The model might resemble a dusty envelope or
a huge spherical atmosphere of the MO. It would appear that this model best fits
the 800 day feature.

\item
Model (C):
This is an initially planar, ring-like structure around the MO.
Even a small amount of angular momentum present in the dust cloud
often tends to form such a structure. We inspected a few different
values of the inner as well as the outer radius of the ring $-$ see
Table~A.5. The minimum inner radius was assumed to be comparable with
the physical radius of the MO depending on its mass and density.

   In the rectangular coordinate frame $O(xyz)$ with the $x$ axis
aligned to the MO's line of apsides and the $x$$-$$y$ plane to be the
orbital plane of the MO, we situated the ring in the following way:
firstly, the frame was rotated clockwise around the coordinate
$z$-axis by the angle $\vartheta$. This new frame $O(x'y'z')$
 was then rotated, clockwise again, around the $x'$-axis by the
angle $\sigma$. (A small number of values for these angles were
inspected but mostly $\vartheta = 30^{o}$ and $\sigma = 45^{o}$ were
used - see Table~A.5.) We obtained frame $O(x''y''z'')$. The new plane
$x''$$-$$z''$ was chosen to be the plane of the ring. 

   Initial results with the disks spanning the range from $5\,000$ to
$10\, 000\,$km were encouraging. Therefore, we tried to improve these
models assuming a non-uniform radial distribution of the DPs in the
ring. A peak in this distribution centered on the MO-centric distance of
$7\, 500\,$km was assumed, whereby the MO-centric distance of the $j$-th
particle, $r_{j}$, in kilometers was calculated according to the relation
$r_{j} = 7\, 500 \pm 2\, 500 \eta^{s}$, where the sign in pair $\pm$ was
randomly generated and $\eta$ was a random number from the interval
$(0,~1)$. Three values for the $s$ index were considered: 1, 3/2, and 2.

   In constructing the C-type model, we kept in the mind the idea that
the ring could persist more than a single orbital revolution of MO
around the central star. This could be possible if the ring consisted of
larger particles, which would have been a source of smaller particles
causing the occultation. The ring could survive the periastron passage
if it were situated within the corresponding Hill's radius. Therefore,
the extent of the ring in a majority of models satisfies this demand.
The ring in models C20 to C34 spans from the physical surface of
a (spherical) MO to the Hill radius. The radius of the physical surface is
calculated assuming a mean density equal to $2\,000\,$kg$\,$m$^{-3}$.
One exception to the stipulation that the whole ring be situated within the
MO's Hill radius is model C13, where the outer radius of the ring is equal to
approximately two MO Hill radii.
\end{itemize}

   In each type of model, we varied the P-R drag parameter $\beta$.
We also created a small number of models to investigate the effect
of changing the standard orbit of a MO and/or the effect of starting a
dust cloud model at distances other than $r_{o}=5\,$au.
% \LEt{Please verify that I have retained the intended meaning and
% modify accordingly.}

\subsection{Evolution of the cloud and light-curves}

   Next, we followed the evolution of the dust cloud and MO during one
periastron passage and back to a distance $r_{o}$. Integrator RA15
\citep{everhart85} within the MERCURY package, version 6, created by
\cite{chambers99} was used for this purpose. The MERCURY integrator
defines `big' and `small' objects. It calculates for the mutual
gravitational interaction of all big objects (i.e., star with
planets and multiple massive bodies) as well as for their effects upon
the small objects (massless particles). On the contrary, small massless
DPs do not affect the motion of massive bodies or that of one another.

   Besides the gravity of the central star and the MO, the motion of the DPs
is obviously influenced by the radiation of the central star. This
action is known as the Poynting-Robertson (P-R) effect. We consider the
basic components of acceleration due to the P-R drag: radial given by
$a_{r} = \beta (GM_{\star}/r^{2})(1 - 2v_{r}/c)$ and transverse given by
$a_{t} = -\beta (GM_{\star}/r^{2})(v_{t}/c)$, where $\beta$ is the ratio
of the P-R drag and gravitational accelerations of the central star, $G$
is the gravitational constant, $v_{r}$ is the radial component of
the star-centric velocity of the DP, $v_{t}$ is its transverse component laying
in the orbital plane of the DP and oriented in the sense of its motion, and
$c$ is the speed of light. The third, perpendicular component of the P-R
drag acceleration is assumed to be zero. The subroutine calculating the
P-R drag acceleration was added into MERCURY version 6.2.

   We assume that all individual DPs are identical and characterized by
the same value of $\beta$. We concentrate on dust grains with
$0.1 < \beta < 1$ but have also  carried out dozens of models beyond this
range (see the tables in Appendix). We neglect other potential
non-gravitational effects.

   Once we have the location of each individual DP as a function of
time, we can calculate the number of DPs transiting
KIC8462 from the point of view of the observer as a function of time.
DPs in transit are simply those that happen to project onto the stellar
disk at a particular time, that is, their distance from the line of sight
towards the exact center of the stellar disk is smaller than the radius
of this star, $R_{\star}$. We consider the coordinate frame having the
reference plane parallel to the sky and with the axis perpendicular to
this plane oriented outward from the position of the observer. In this
coordinate frame, the inclination of the orbital plane, $i$, of MO is
$i = 90^{o}$ (observer is sitting in the orbital plane of MO). We denote
the angle between the line of apsides and observer's line of sight to
KIC8462, measured in the direction of the MO's motion, by $\Phi$ as
shown in Fig.~\ref{scheme}. In this figure, the crossing of the reference
plane and orbital plane of MO is shown with the dashed line. Angle
$\Phi$ is related to the argument of periastron of MO, $\omega$, via
$\Phi + \omega = 270^{o}$.

   We assume the set of values of angle $\Phi$ ranging from $0^{o}$ to
$360^{o}$ with increments of $1^{o}$ but we exclude the interval
$170^{o}$$-$$190^{o}$, which is near the apastron. We consider a cloud
of $1\,000$ DPs in a broad parameter space but use $10\,000$ particles
when zooming into a few specific locations in the grid of our
A-models (see Tables A.1$-$A.3). Finally, we convert the number of
particles in transit, $n(t)$, into a synthetic light-curve, $f(t)$, for
comparison with the observations, by shifting it to the proper BKJD
values and scaling its intensity as follows:
\begin{equation}
f(t) = 1 - n(t - t_{m} + t_{o}) \frac{a_{o}}{a_{m}}, 
\end{equation}
where $t_{m}, a_{m}$ are the time and amplitude of the model ($n(t)$
function) while $t_{o}$ and $a_{o}$ are the time and depth of the
observed feature. This simple and fast procedure is equivalent to the
assumption that the dust is optically thin and that it only absorbs and
scatters the radiation out of the beam while the scattering emission
into the line of sight is ignored. The limb darkening is not taken into
account.

\section{Results}

   We calculated the synthetic light-curves in 187 models of type
A, 27 models of type B, and 39 models of type C. The initial parameters
of all models are given in Tables~A.1 through A.5 in the Appendix.
Although a quantitative match between the model and theory with only a
small number of free parameters is very hard to achieve, a surprising
qualitative, morphological similarity appeared for several models, the
best of which we describe in the following.

   To evaluate the match between the observed event and
corresponding model, we attempted to find some automatic minimization
technique that would pick the best matching models from the set of
created models but were not successful. This is most probably
due to the fact that the observed features are too complicated (often
with multiple peaks), the range of free parameters is too large (many
orders of magnitude), while our coverage of the parameter space is very
limited (with large steps). For, this reason, we manually reviewed all the
synthetic light-curves. In spite of such a subjective method, however,
we hope to demonstrate that almost all observed features can be
understood within an appropriate simple model.

\subsection{Feature at 800 days}

\begin{figure*}
\centerline{\includegraphics[height=8.cm, angle=-90]{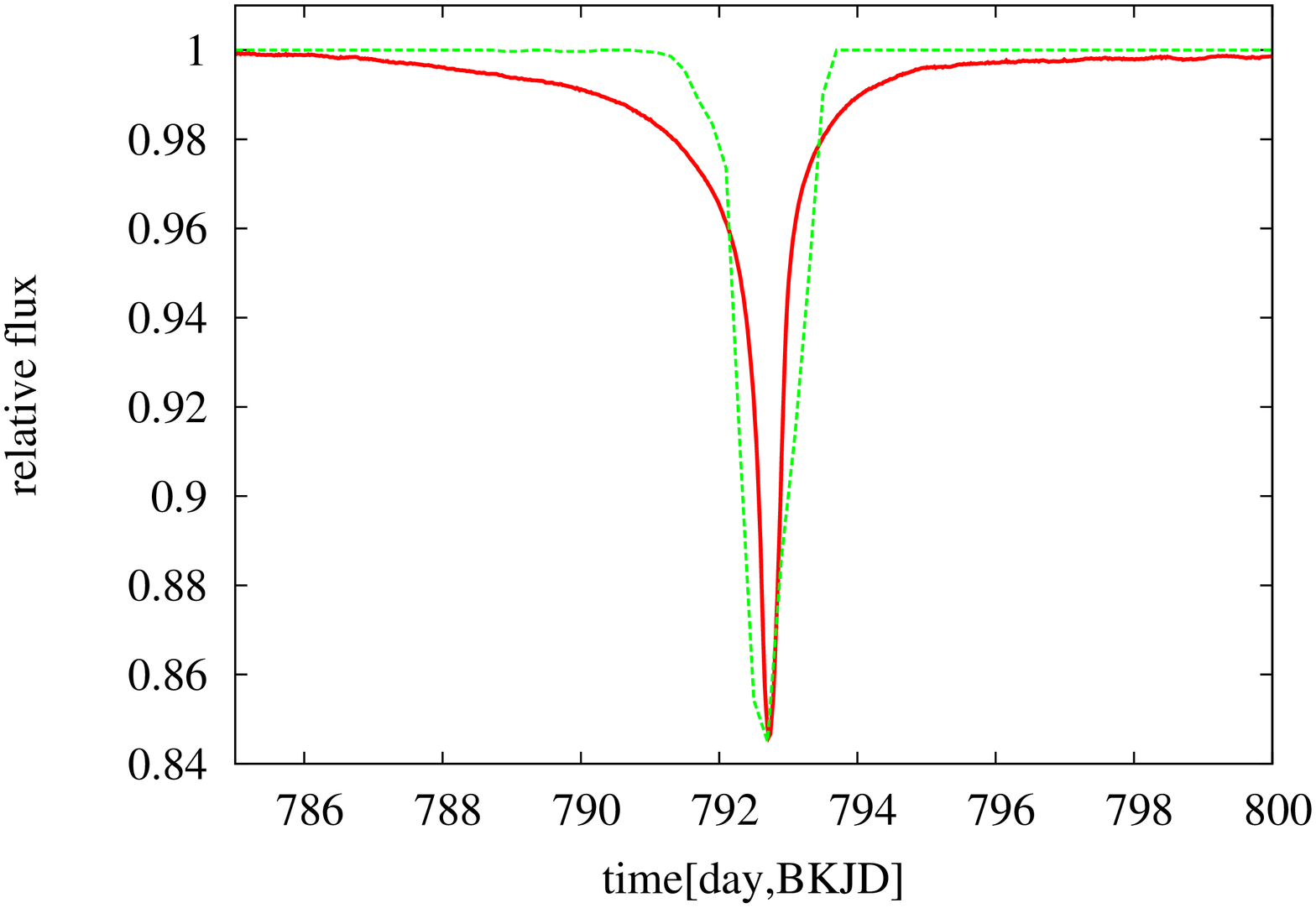}
            \includegraphics[height=8.cm, angle=-90]{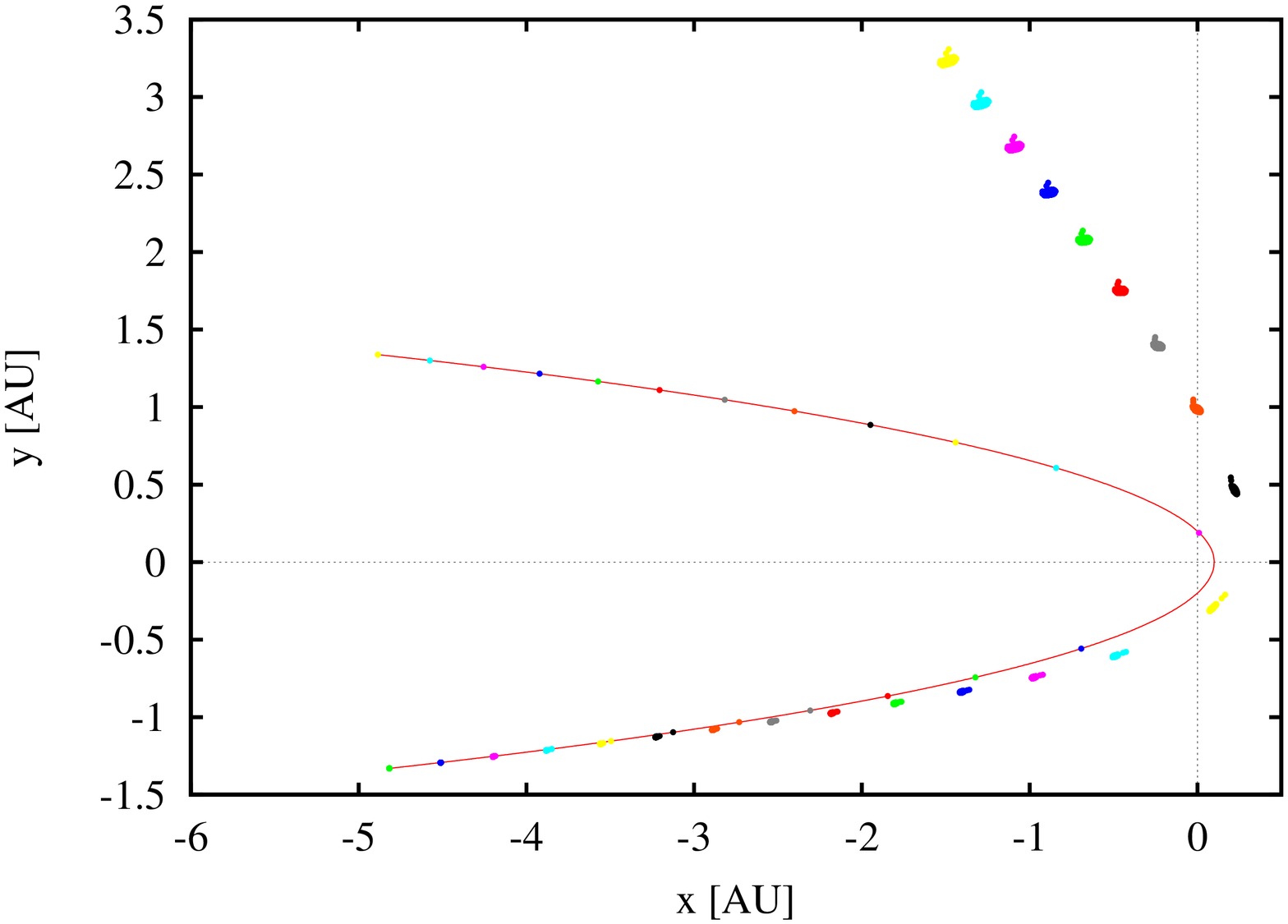}}
\centerline{\includegraphics[height=8.cm, angle=-90]{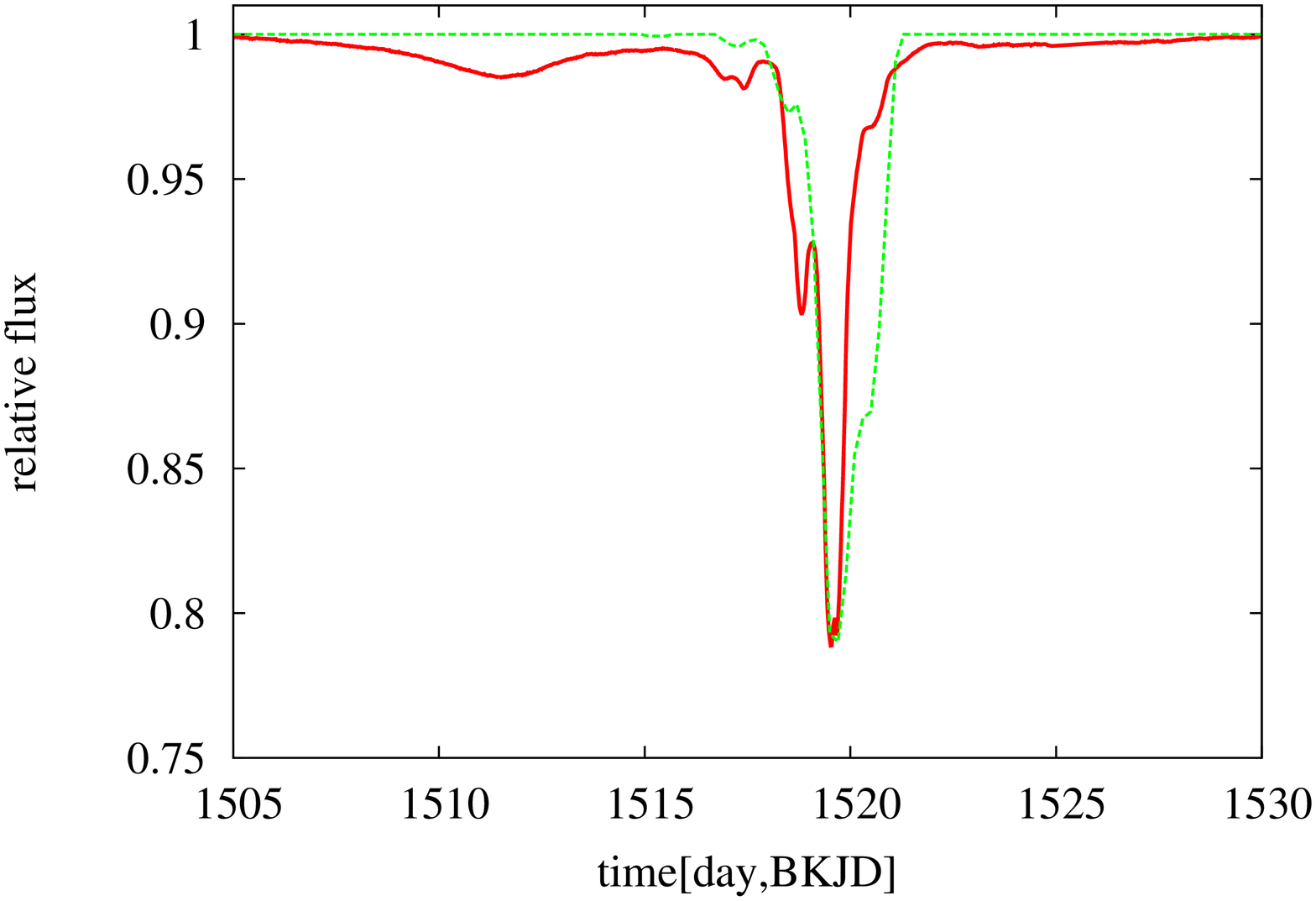}
            \includegraphics[height=8.cm, angle=-90]{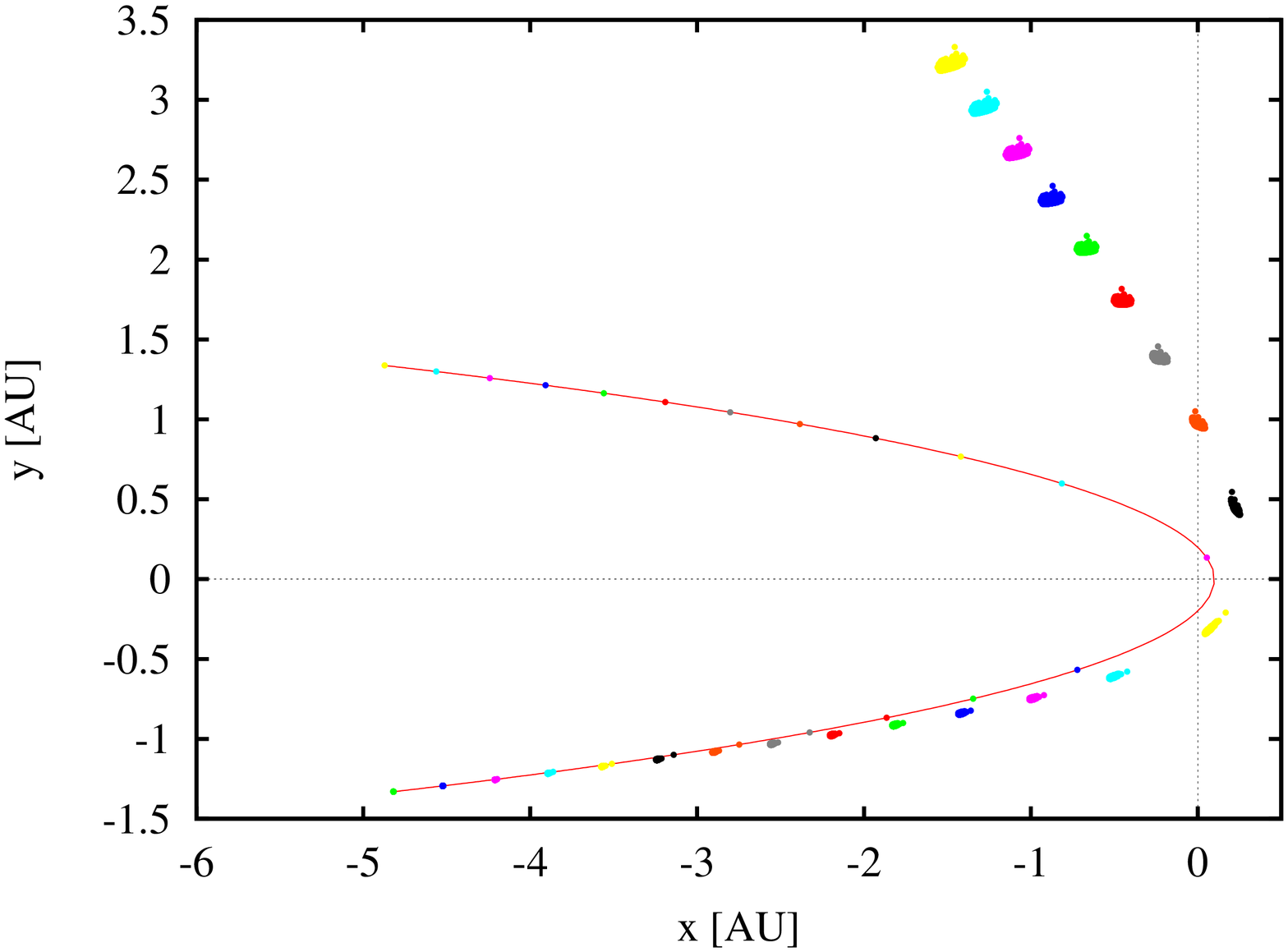}}
\centerline{\includegraphics[height=8.cm, angle=-90]{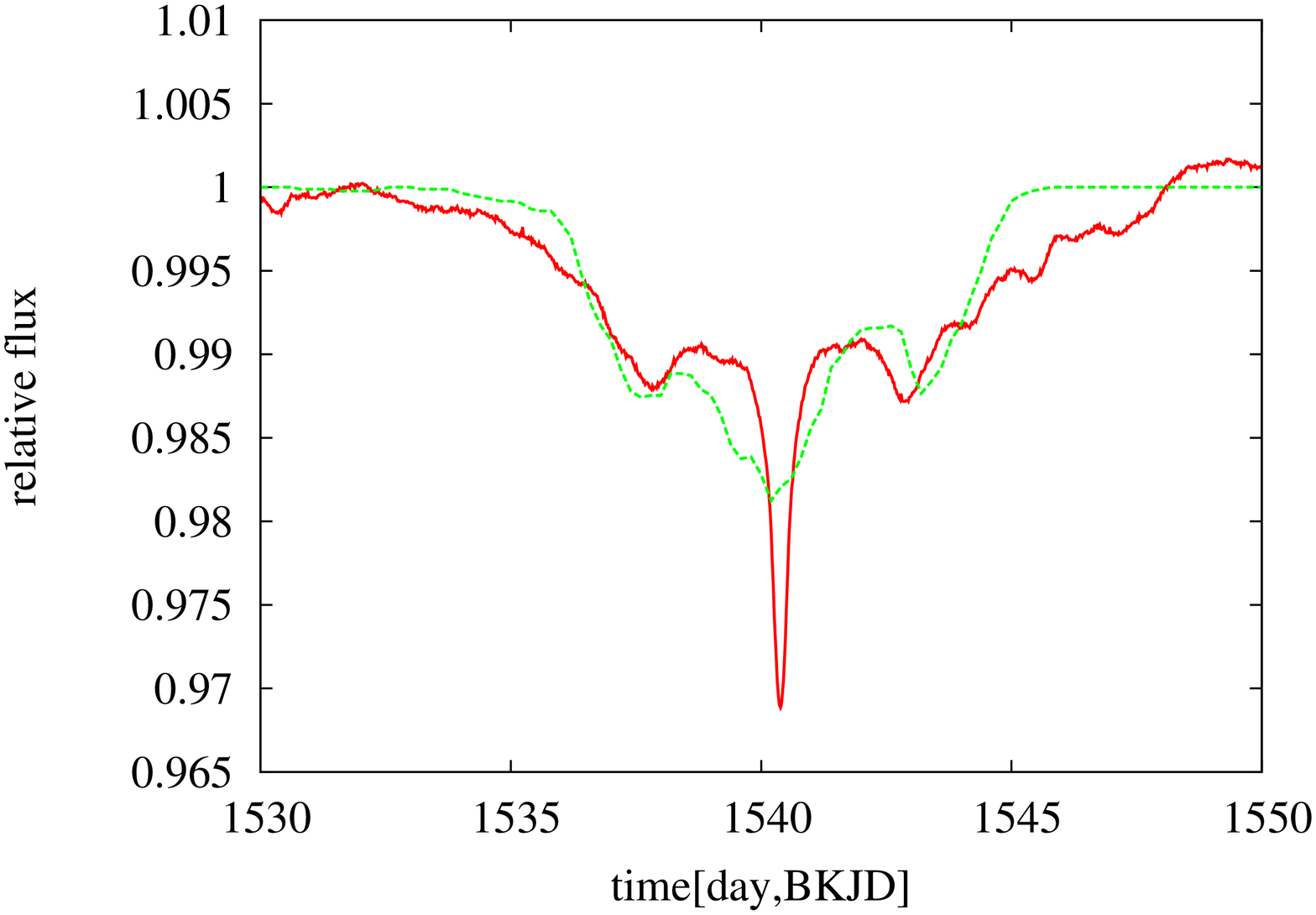}
            \includegraphics[height=8.cm, angle=-90]{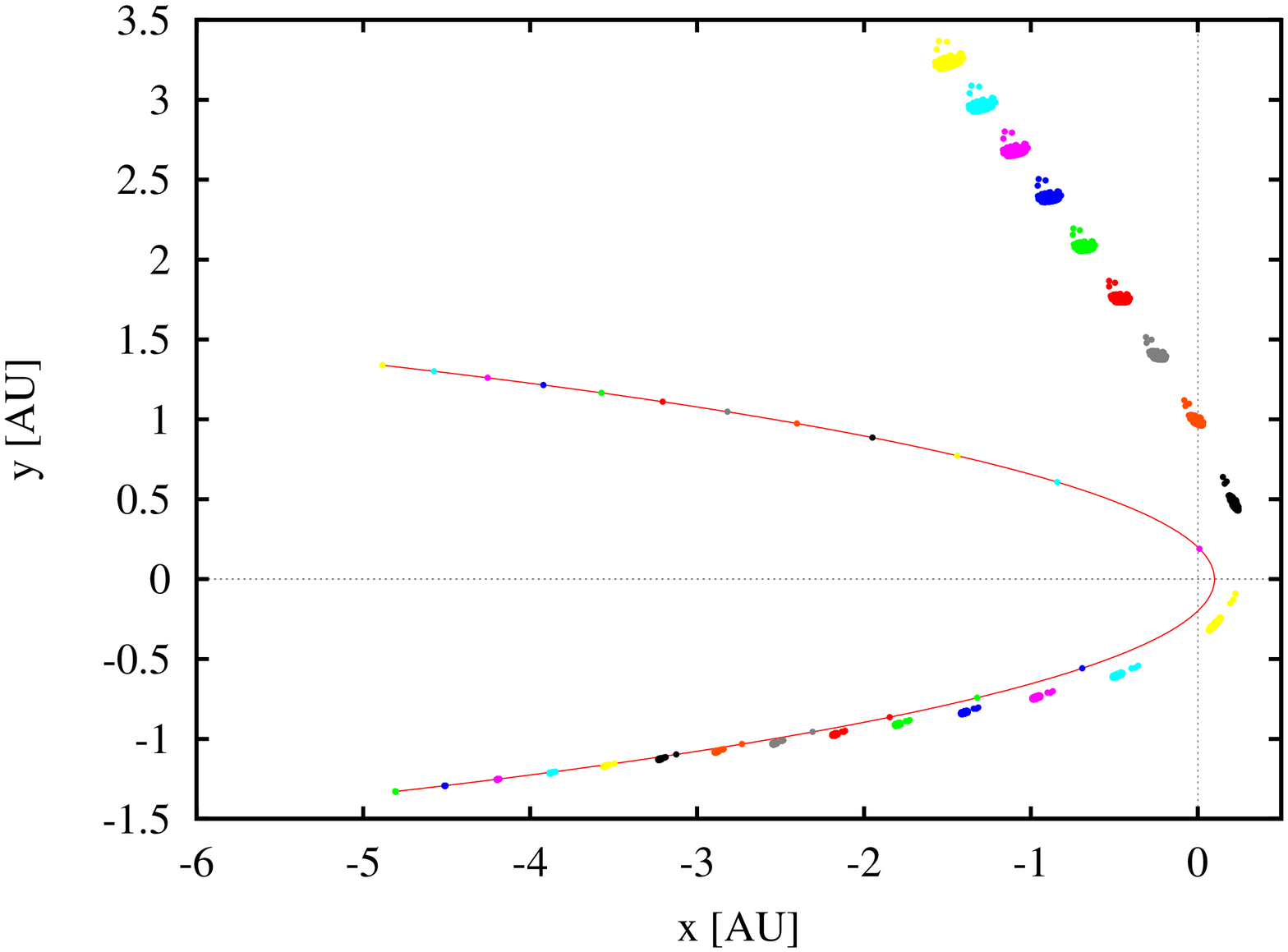}}
\centerline{\includegraphics[height=8.cm, angle=-90]{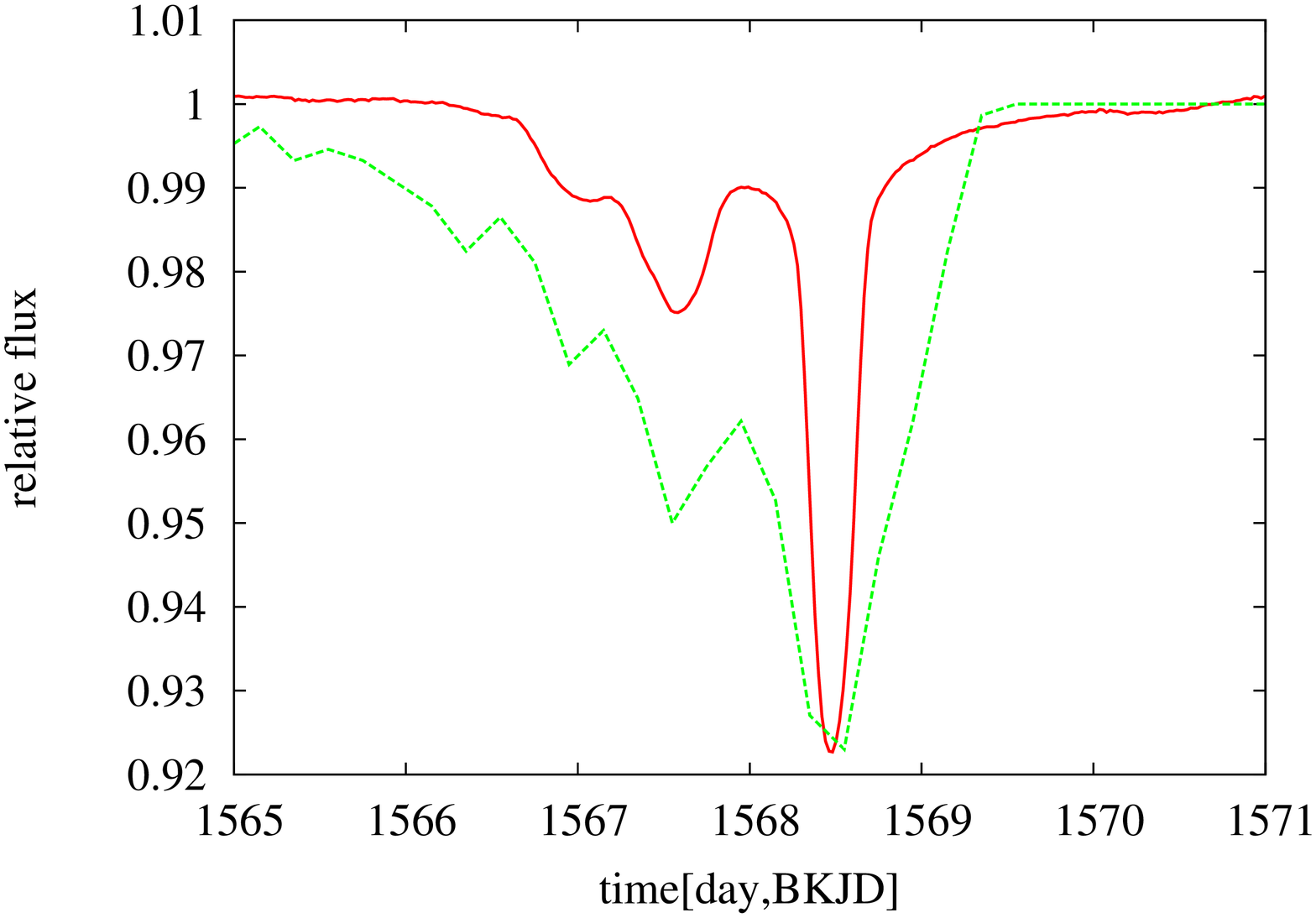}
            \includegraphics[height=8.cm, angle=-90]{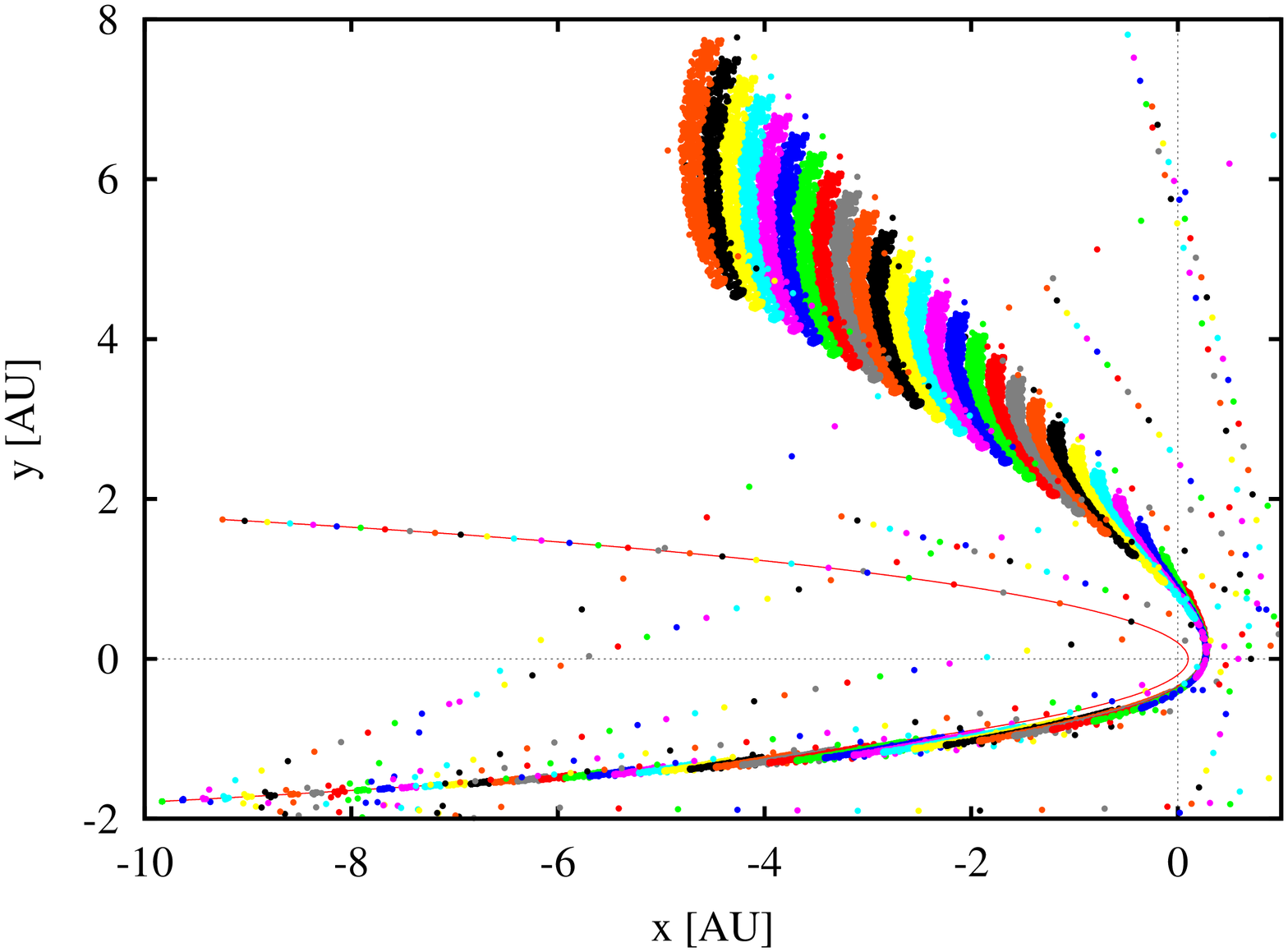}}
%\centerline{\includegraphics[width=8.5cm, angle=0]{fit0800.pdf}
%            \includegraphics[width=8.5cm, angle=0]{cloud0800.pdf}}
%\centerline{\includegraphics[width=8.5cm, angle=0]{fit1520.pdf}
%            \includegraphics[width=8.5cm, angle=0]{cloud1520.pdf}}
\caption{Comparison of observations with model (A). Each row
corresponds to one of the 800, 1$\,$520, 1$\,$540, and 1$\,$570 day
features. Left column: the observations (red/solid line) and model (A)
(green/dashed). The fits may not be perfect but the main morphological
features are represented. Right column: the cloud of particles
(dots) and its parent body (solid line) orbiting the star. Different 
colors are used to plot particles at different times separated by 25 day
intervals. A small number of particles may suffer from close 
encounters and were thrown into chaotic orbits.}
\label{modela}
\end{figure*}

   This feature is a single minimum with a smooth decline and a sharper
egress. It was not explained in terms of the comet scenario. This kind
of shape is produced naturally in our models since as the dust cloud
approaches the periastron, it shrinks and creates a sort of leading tail.
Fig.~\ref{modela} displays one of them and how it compares to the
observations. This particular model was type (A) and was obtained
assuming % q01qtp75Qtp1000m11B0.9phi029
mass, periastron, and apastron of the MO: $10^{-11}\,$M$_{\star}$,
$0.1\,$au, and $50\,$au, respectively (model A39 in Table~A.1).
The spherical dust cloud was composed of DPs having the properties   
corresponding to $\beta=0.629$, which were initially placed on
elliptical orbits around MO with pericenters and apocenters of $75\,$km
and $75\, 000\,$km. Start/end of integration was at the pre-periastron
star-centric distance of MO equal to $5\,$au and the line of sight had
$\Phi=29^{o}$.

\begin{figure*}
\centerline{\includegraphics[height=8.5cm, angle=-90]{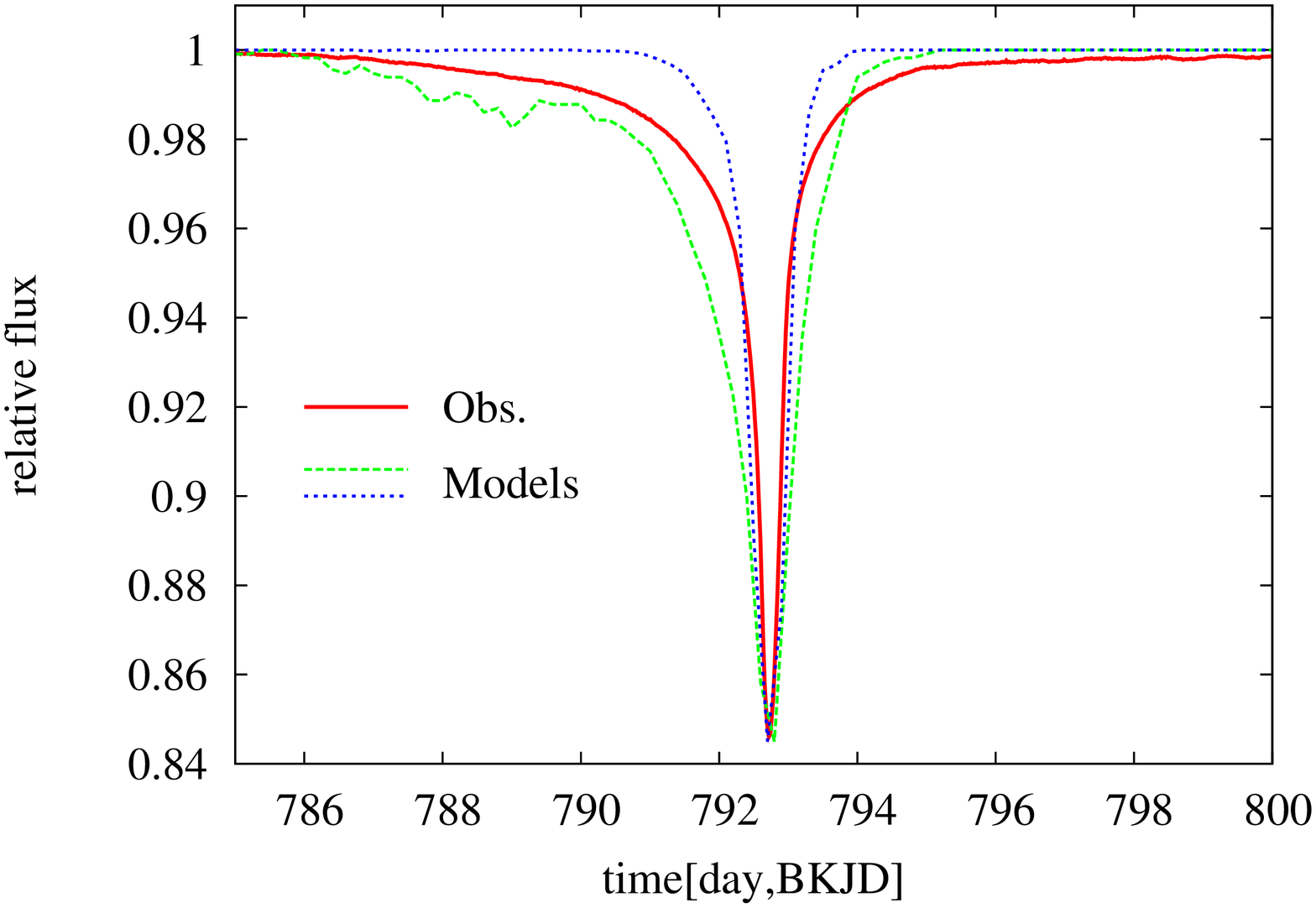}
            \includegraphics[height=8.5cm, angle=-90]{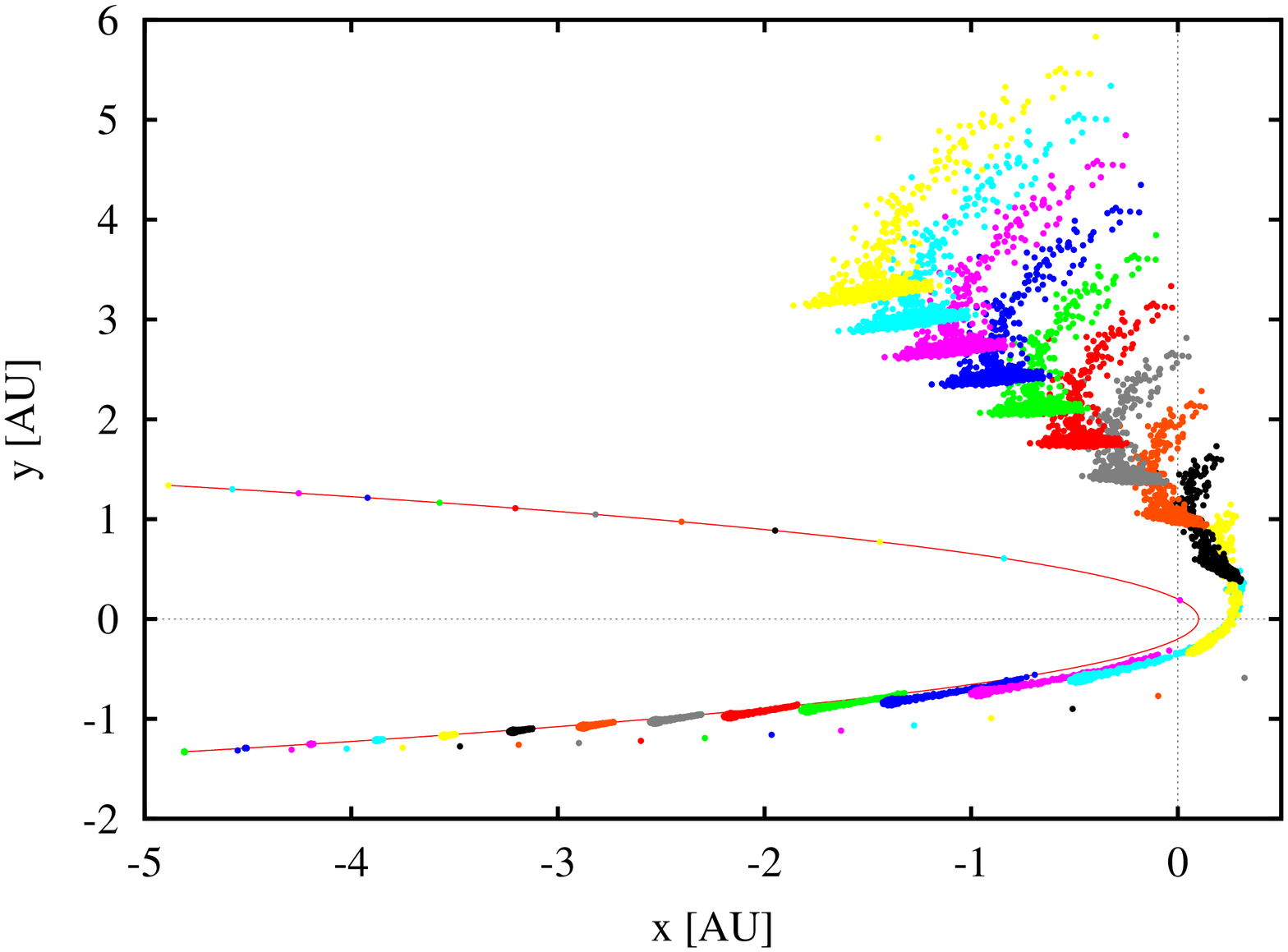}}
\caption{Comparison of observations with model (B). Left column: the
observations (solid line), two models (green-dashed and blue-dashed;
see the text for a description). Right column: the cloud of particles
and its parent body corresponding to the green-dashed model orbiting 
the star at 25-day intervals distinguished by different colors.}
\label{modelb}
\end{figure*}

   However, it appears that the models of type (B) can reproduce this kind
of shape even better. % GRID2m1m10B0.9phi029 % GRIDn2m1m8B0.9a3phi336
Two of them are depicted in Fig.~\ref{modelb} and their parameters
are listed in Table~A.4 $-$ models B18 and B5. The periastron and
apastron of the MO are the same as in the A39 model. The dotted
blue line is for the MO mass of $10^{-10}\,$M$_{\star}$ (B18). 
The DPs constituting the spherical dust cloud move in orbits with pericenters equal to $1\,000\,$km. The interval of the randomly
distributed initial velocities in the pericenter corresponds to the
interval of apocenters randomly distributed from 10 to 100 times of
the pericenter distance. The same good fit can be obtained for the same value
of angle $\Phi = 29^{o}$.

   The dashed green line corresponds to the MO mass of $10^{-8}\,$M$_{\star}$ (B5).
The DPs constituting the spherical dust cloud initially move in orbits
with pericenters equal to $1\,000\,$km. The interval of the randomly
distributed initial velocities in the pericenter corresponds to the interval
of apocenters randomly distributed from 30 to 300 times the pericenter
distance in this case. The good fits are observed also at the pre-periastron
part of the orbit and this particular one is for angle $\Phi = -24^{o}$.
The properties of the DPs were identical to those in model (A), that is,
$\beta=0.629$.

% FIG. 5

% FIG. 6

% FIG. 7
\begin{figure*}
\centerline{\includegraphics[height=8.5cm, angle=-90]{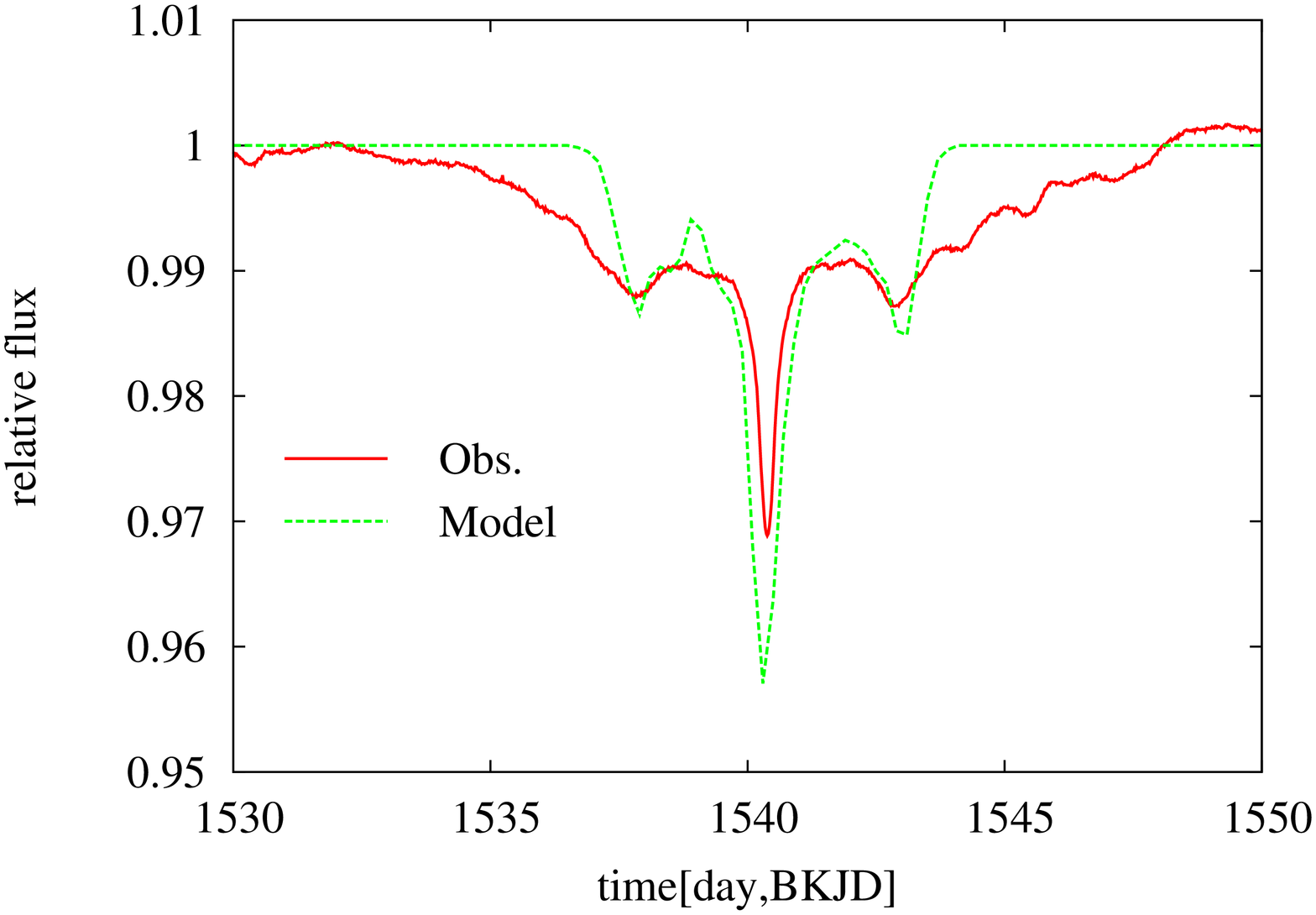}
            \includegraphics[height=8.5cm, angle=-90]{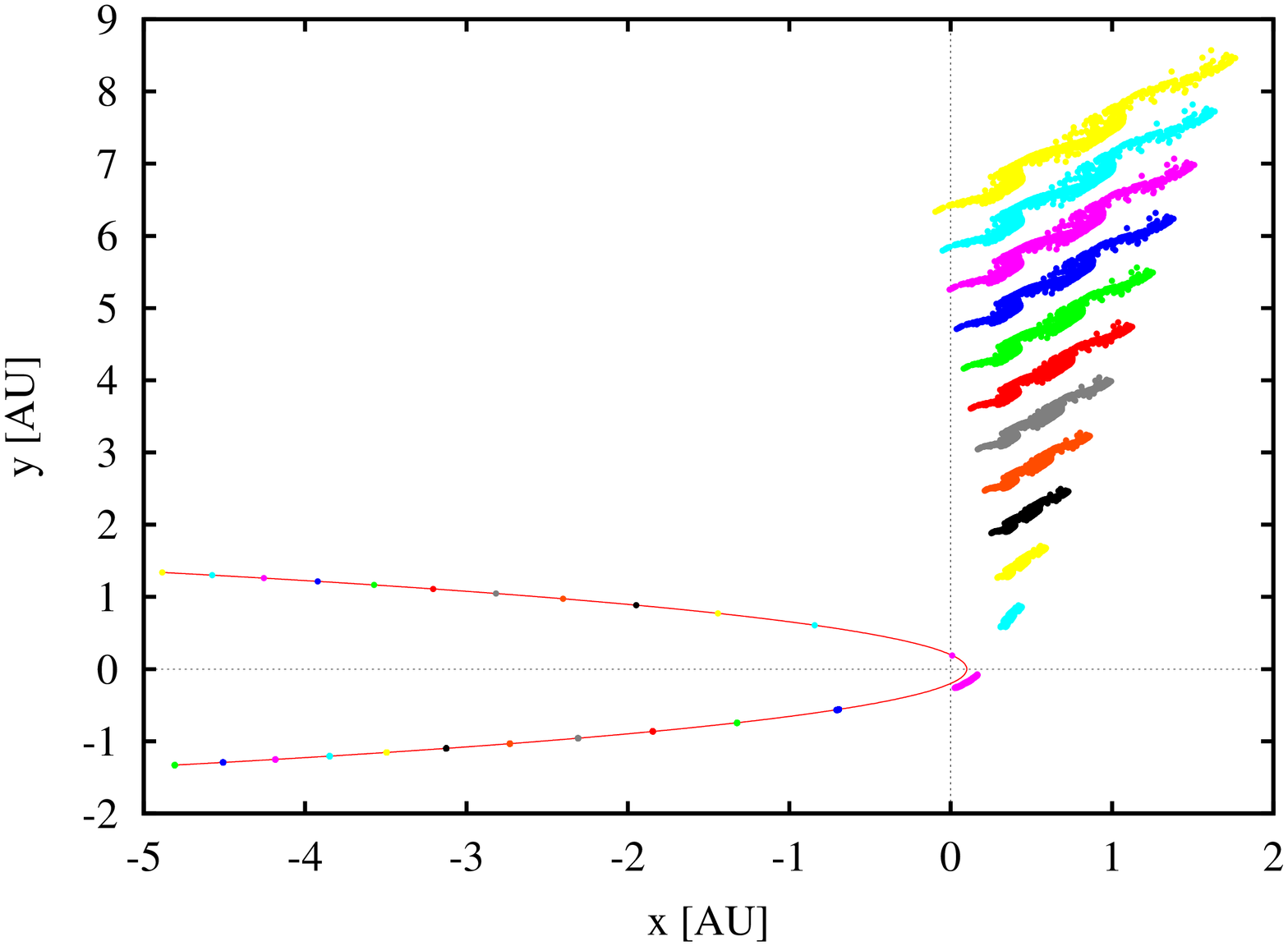}}
\caption{Comparison of observations with model (C). Left column: the
observations (solid line), model (dashed). Right column: the cloud of
particles and their parent body orbiting the star at 25-day intervals.}
\label{modelc}
\end{figure*}

\subsection{Feature at 1$\,$520 days}

   This is the deepest feature and is very complex. It contains several
peaks, which gradually increase in depth before the global minimum is
reached, and there is one further bump on the egress. This latter feature may be
analogous to that at 800 days in the sense that if it were smoothed
by some effect, the two would resemble one another. Again, we were surprised to
see a number of models, with only one MO and a simple dust cloud, which
were able to reproduce the essence of this complicated morphology. One
of them is illustrated in Fig.~\ref{modela} in the second row from the
top. This particular model is also A-type (A4) % GRID1m1m10B0.9
and was obtained with the following assumptions, which are very similar
to the previous model. Mass, periastron, and apastron of the MO:
$10^{-10}\,$M$_{\star}$, $0.1\,$au and $50\,$au, respectively. The
spherical dust cloud composed of DPs with $\beta=0.629$ placed initially
on elliptical orbits around MO with pericenters and apocenters of
$1\,000\,$km and $100\, 000\,$km. Start/end of integration was at a
distance of MO equal to $5\,$au and the line of sight had $\Phi=29^{o}$.
The model fits the location of almost all bumps and, qualitatively, also
agrees with their strengths. 

   Notice that there is a shallow and smooth bump at BKJD 1511
which was not reproduced with such a model.
One could speculate that an additional body/cloud is required to bring
about the above observations or that they could be caused by, for
example, a population of particles on slightly different orbits with
slightly lower values of $\beta \approx 0.559,$ which would allow them
to move slightly faster on the pre-periastron part of the orbit. The same
values of periastron, apastron, and angle $\Phi$ implies that both
objects and associated dust clouds responsible for the 800 and 1$\,$520
day features, moved in very similar orbits, with the same orientation in
space and also suggests that the two objects might have a common
progenitor.

\subsection{Feature at 1$\,$540 days}

   This feature has three main peaks, the middle one being the
strongest. The feature differs from the others since it appears
symmetrical, thus invoking ideas of a body with a ring structure
transiting the star. Also, this structure is in agreement with our
calculations. We observed numerous instances of triple-peak structures
with peaks moving in time and intensity for model (A).
The third pair of panels in Fig.~\ref{modela}
% \LEt{I can't distinguish what is being refezrred to hear as
% ``the middle'' in figure 5. The author may want to consider referring
% to a particular panel or feature of this figure to improve clarity.}
shows one example of the A-model (A27) applied to this feature. % q01qtp100Qtp1000m10B0.9
We were not able to fit the width of the main peak completely, but nevertheless, from rather high angles (e.g., $\Phi=94^{o}$), the
calculations grasp the main morphological structure.
The following parameters were used: mass, periastron, and
apastron of the MO: $10^{-10}\,$M$_{\star}$, $0.1\,$au, and $50\,$au,
respectively. The spherical dust cloud composed of DPs with
$\beta = 0.629$ initially placed on elliptical orbits around MO with
pericenters and apocenters of $100\,$km and $100\,000\,$km, respectively. Start/end of
integration was at a distance of MO equal to $5\,$au and the line of
sight had $\Phi=94^{o}$. Thus, this might be a considerable smaller
body compared to the two MOs mentioned above, but on almost the same
orbit, only its argument of periastron would have to be tilted to
account for different line of sight angle.

   We also found that some models of type (C) fit the feature quite
well even if we assume an initially uniform distribution of the DPs in
the ring. A significantly better match was obtained, however, when we
considered the ring of the DPs with the distribution peaked in the
middle. The best model for the 1540-day structure is illustrated in
Fig.~\ref{modelc} and its parameters are given in Table~A.5, model C37.
% GRID10...; model C6.} % GRID6m1m8B0.9/
The position and width of the three peaks fit very well and only the
intensity of the central peak is stronger. The observed feature
also exhibits broader outer wings. This fit was obtained with the
following parameters. It is an inclined ring with $\vartheta = 30^{o}$
and $\sigma = 45^{o}$, inner and outer radius of $5\,000$ and
$10\,000\,$km, respectively. The initial distribution of the DPs in
the ring was generated using the formula for the calculation of
MO-centric distance of $j$-th particle
$r_{j} = 7\,500 \pm 2\,500\eta^{3/2}$.
The mass of the central body was $10^{-8}\,$M$_{\star}$, and its orbit
was the same as before. The line of sight had $\Phi=29^{o}$.
Notice, that the ring breaks and decouples from the object after 
passing the periastron.

\begin{figure}
\centerline{\includegraphics[height=8.5cm, angle=-90]{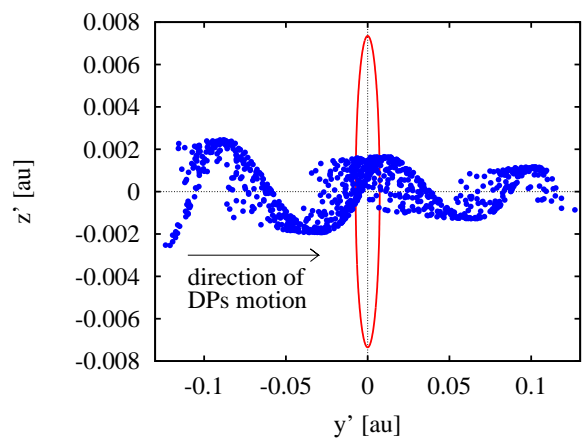}}
\caption{Dust cloud with individual particles (dots) corresponding to
the C-model (C37) passing in front of the star (ellipse) during
the eclipse causing the observed $1\,540$ day feature.}
\label{ring}
\end{figure}

   The radiative acceleration causes periodic ripples in the cloud,
which in 3D resemble a sort of squeezed spiral, and are the reason for
the observed light-curve structure. This is illustrated in
Fig.~\ref{ring}, which displays the individual DPs passing in front of
the star at a moment during the eclipse. So, in this model, the three
main peaks in the light-curve are not due to a ring passing in front of
a star, although there was a ring structure before. It is very
encouraging that this parent body has the same orbit as the two bodies
before and even the line of sight is the same. For this reason we
prefer the model (C) rather than the model (A) for this object.

% FIG. 8

\subsection{Feature at 1$\,$570 days}

   This feature exhibits three dips increasing in depth,
which resembles the feature at $1\,520$ days or the gradual ingress of
the feature at 800 days. Our model is displayed in Fig.~\ref{modela} in
the bottom part. It fits very well the position of all peaks but the
strongest one is more shallow and wider than expected. It was obtained
using the following parameters (model A124 in Table~A.3).
% R10q01Q10qm1m10B0.9/; r_o = 10 AU
Mass, periastron, and apastron of the MO: $10^{-10}\,$M$_{\star}$,
$0.1\,$au, and $50\,$au, respectively. Spherical dust cloud is composed
of DPs with $\beta = 0.629$ initially placed on elliptical orbits around
MO with pericenters and apocenters of $1\,000\,$km and $10\, 000\,$km, respectively.
This time, start/end of integration were at a distance of MO equal to
$10\,$au and the line of sight had $\Phi=0^{o}$. This body and orbit are
also very similar to those responsible for other features and only the line
of sight (i.e., the argument of periastron) is shifted.

\section{Discussion, comments, and speculations}
\label{discuss}

   As we have already stated, the question of the origin or the long-term stability of dust clouds we modeled are beyond the scope of
the present paper. Nevertheless, these are important questions that can
significantly affect the probability of observing the studied
events. That is why we present a number of tests, comments and/or 
speculations to address this issue.

   Clouds composed of particles with high $\beta$ values are extremely
vulnerable and easily decouple from the parent massive body and will not
be observed during multiple transits. This means that the chance of observing such events is rare compared to low-$\beta$ particles. On the
other hand, large particles with low $\beta$ values might stay within a
Hill radius of the parent body and return back to cause new eclipses
provided that they are not perturbed by other MOs. It might be that they
serve as a dust reservoir and, upon their return, collide and produce
new smaller and more opaque debris particles that can obscure the star
more easily.

   The Hill radius $R_{H}$ for an MO on a
highly eccentric orbit with eccentricity $e$ and semi-major axis $a $ is based on the calculations of \cite{hamilton92} and is analogous to the
common Hill radius for a circular orbit, however it assumes the pericenter
distance rather than the radius, that is,
\begin{equation}
R_{H}= \left( \frac{m}{3M_{\star}}\right)^{1/3} a(1-e).
\end{equation}

\subsection{Perturbation of the dust cloud near the periastron}
\label{6.1}

   A scenario that we have not addressed so far is a possible interaction
between the massive objects and their dust clouds. In this work, the
clouds were studied during only a fraction of their orbit in the
star-centric distance typically equal to or less than $5\,$au.
% from their parent MO\LEt{Please check that I have retained your
% intended meaning.}.
Although they were modeled independently, their parent
MOs may have a common origin, they may be in relatively close proximity to one another, and may therefore affect
each other as well as their dust clouds. A feature observed at 800 days
is separated by 730 days from the other three features and by approximately 530
days from the two very small features observed in the light-curve earlier
(they are too small and are not modeled in this paper). However, the
latter three objects (features) are separated only by
$\Delta t \approx 20$ days in the time domain. Assuming that they follow
one another on the same orbit, this time lag can be translated into
their physical separation $\Delta l$ , which will decrease with their
distance from the star as, approximately, 
\begin{equation}
\Delta l \approx v \Delta t \approx 
\sqrt{GM_{\star} \left( \frac{2}{r}-\frac{1}{a}\right)} \Delta t.
\label{dl}
\end{equation}

   To investigate a mutual influence of the two MOs on one another as
well as an influence of the second MO on the dust cloud of the first MO,
we carried out the following simulation. We assumed a MO orbiting a
star with the same orbit as before, that is, mass, periastron, and apastron
$10^{-8}\,$M$_{\star}$, $0.1$, and $50\,$au, respectively. Further, we
assumed the dust cloud of type (A) of massless particles placed
initially on orbits around the MO as before. The orbits of the DPs were
oriented randomly and were all given pericenters and apocenters equal to
$1\,000\,$km and $100\, 000\,$km, respectively. Since we intend to test
the stability of the cloud in respect to the gravitational perturbations,
no P-R drag was considered. The motion of the MO and its cloud was
integrated during a fraction of the orbital period at star-centric of
less than $5\,$au.

   In the following step, we repeated the same integration but with another MO
added to the system. The latter was placed in an identical orbit to the
first MO but with a time lag of 20 days as indicated by the
observations.
The mass of the second MO was chosen to be $10^{-8}\,$M$_{\star}$ to
maximize its perturbation effect. Then we compared the results of the
calculations with and without the second MO. Even in this extreme case,
the change of the orbit of the first MOs as well as the structure of its
dust cloud was negligible. We conclude that our MOs and their dust
clouds can be regarded as independent entities during the fraction of
their orbit within the star-centric distance of $5\,$au considered in
this paper and where the obscuration events have happened. This a
posteriori justifies our presumption in the calculations and choice to model only one MO and one dust cloud at a time.

\subsection{Perturbations between the massive bodies}
\label{twoMOs}

   After the two MOs, following each other with a time lag of 20 days on
the same orbit passed the periastron, the physical distance between them becomes shorter
according to Equation \ref{dl}, implying that the first MO moves
slower than the second. At some point, they may become too close and
their mutual gravity might kick in. This would shrink the distance between
them even closer and a strong mutual interaction might occur affecting
their dust clouds. To investigate what might happen beyond that part of
the orbit considered in our previous calculations, we explored in the
following scenario.

   We assumed a MO orbiting the KIC8462 with the same orbit as before,
that is, periastron and apastron of $0.1$ and $50\,$au, respectively, and a mass
of $10^{-8}\,$M$_{\star}$. In the beginning of our simulation, this MO
was situated in the pre-periastron arc at a star-centric distance of
$5\,$au. Another MO in the same orbit with the same mass followed
the first MO with a time lag $\Delta t = 20$ days. We integrated this
system over a single orbital period of the MOs on their initial
Keplerian orbit (104 years). It indeed appears that both MOs influence
each other during a short period of mutual close approach, which
occurred $\sim$$7\,530$ days after the first MO passed the periastron.
In the post-periastron arc of orbit, the second MO approaches the first
MO since its star-centric speed is larger. Its distance to the first MO
is also reduced by the mutual gravity of the MOs.

\begin{figure*}
\centerline{\includegraphics[height=8.5cm, angle=-90]{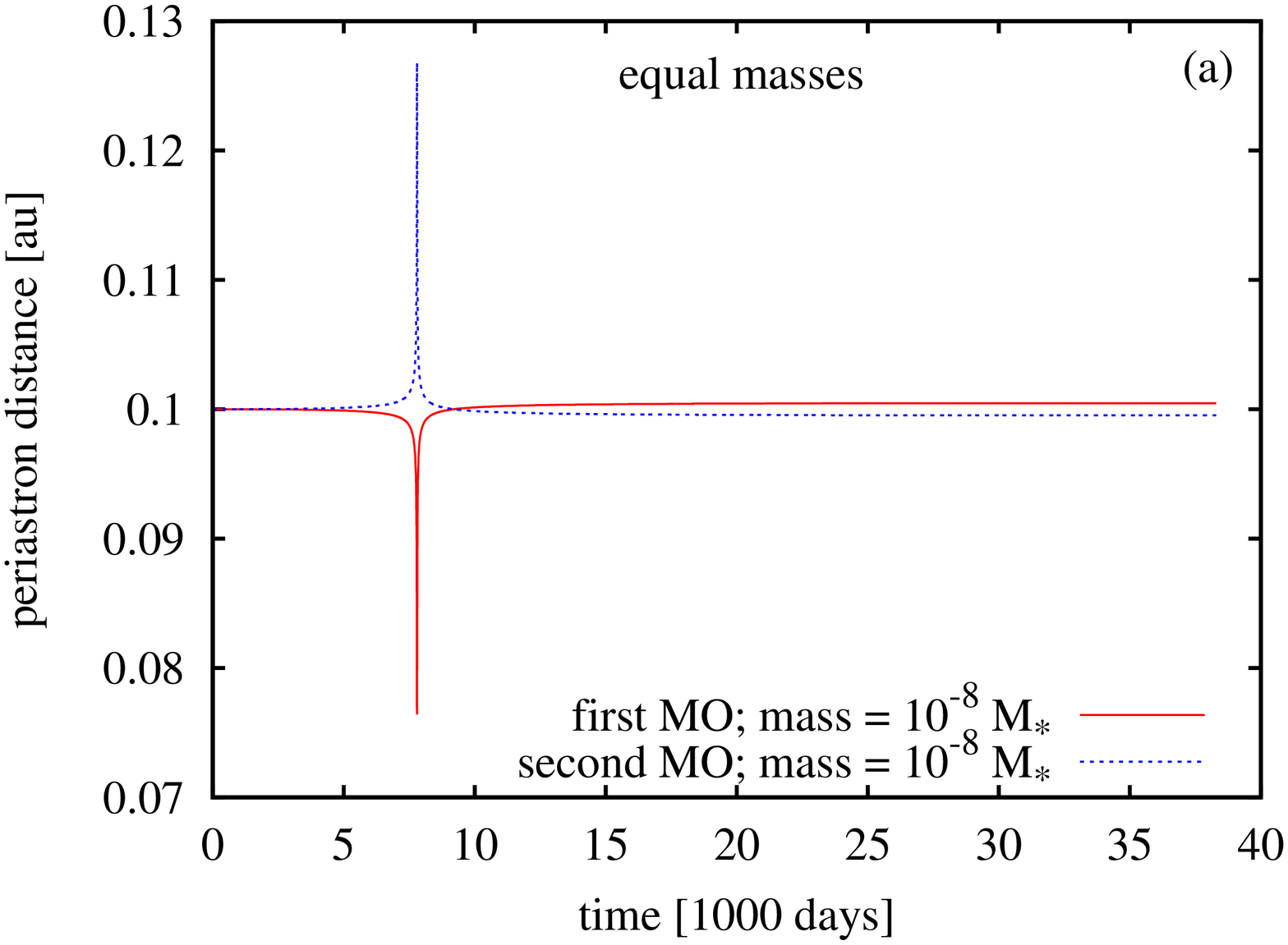}
            \includegraphics[height=8.5cm, angle=-90]{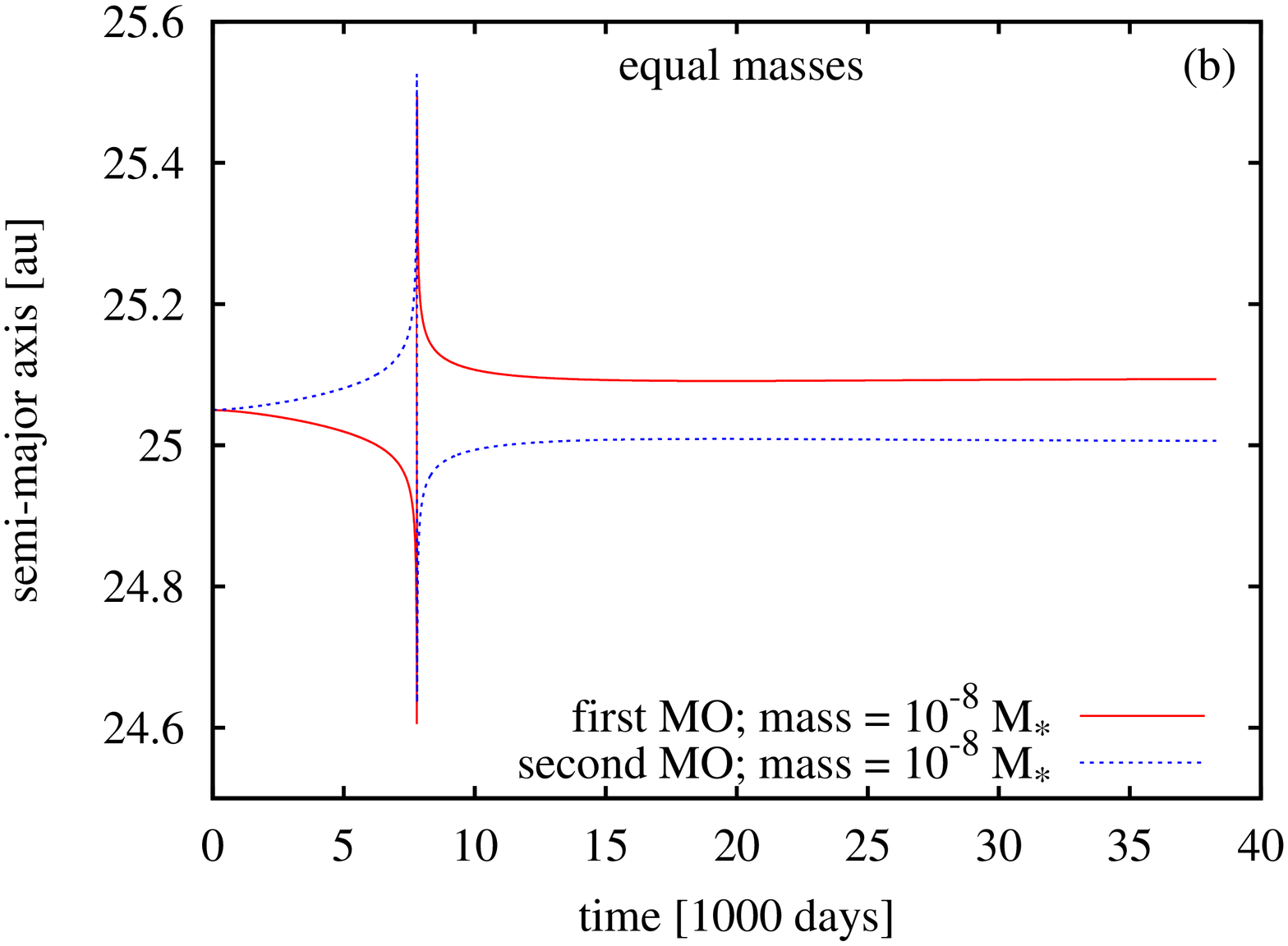}}
\centerline{\includegraphics[height=8.5cm, angle=-90]{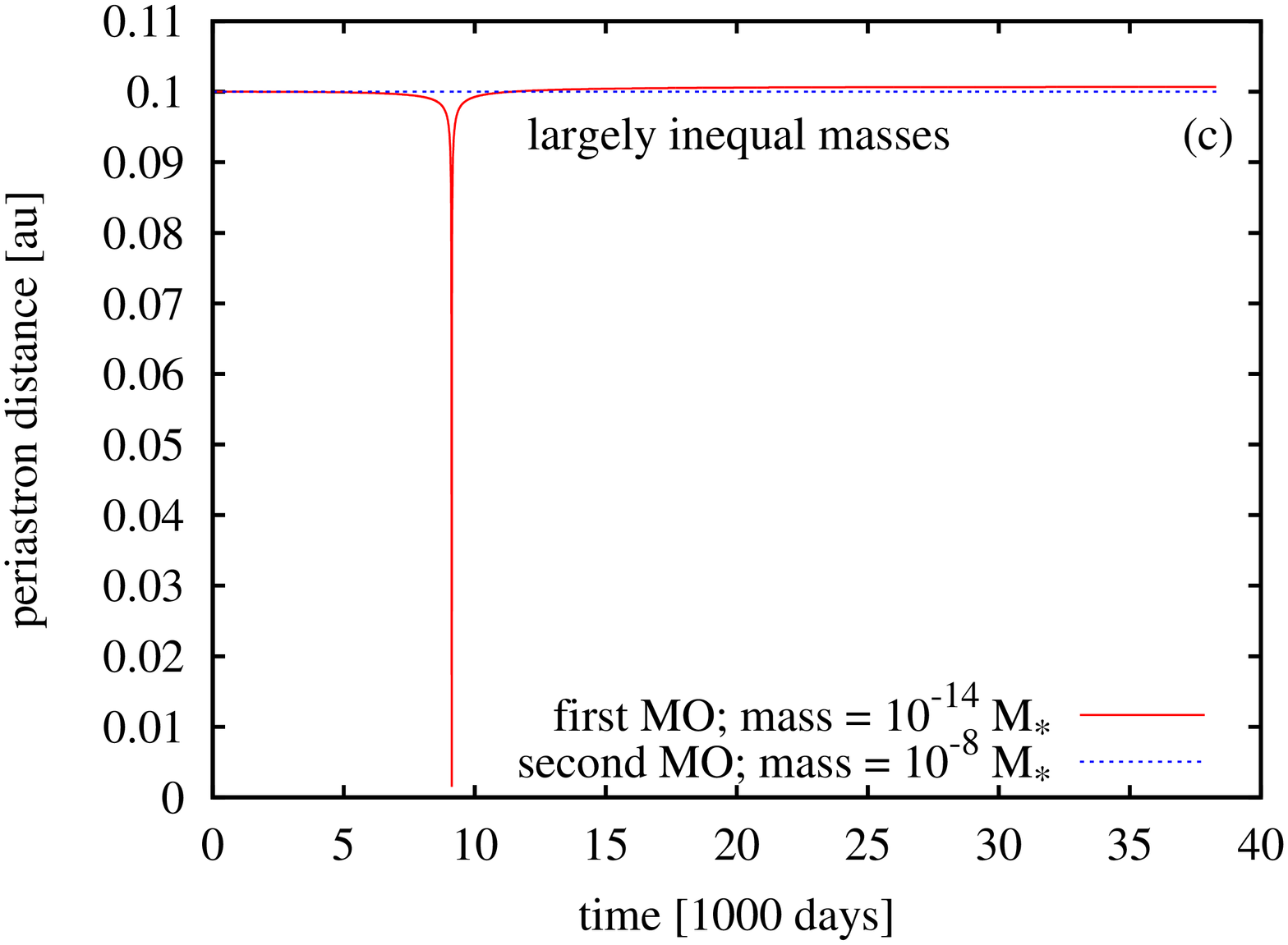}
            \includegraphics[height=8.5cm, angle=-90]{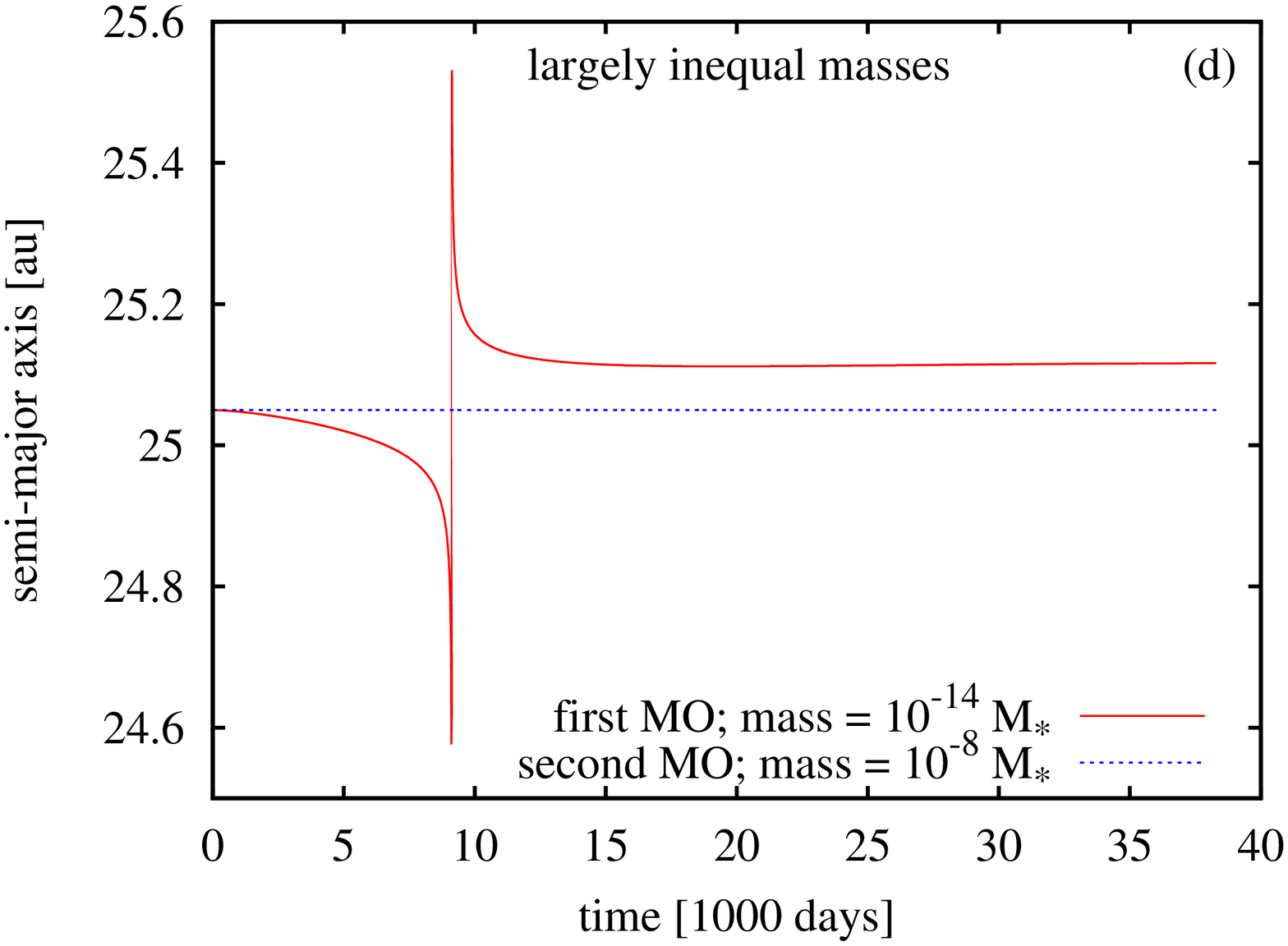}}
\caption{The evolution of periastron distance (plots a and c) and
semi-major axis (plots b and d) of two mutually perturbing MOs moving
around the KIC8462 in orbits described in Sect.~\ref{discuss}. In
the first case (plots a and b), the masses of both MOs are the same
and equal to $10^{-8}\,$M$_{\star}$. In the second case (plots c and d),
the mass of the first (second) MO is $10^{-14}\,$M$_{\star}$
($10^{-8}\,$M$_{\star}$).}
\label{pertMOs}
\end{figure*}

   After $7\,530$ days from the moment when the first MO passed the
periastron, the MOs are separated by only approximately $0.00082\,$au and a
strong perturbation occurs. The change of periastron distance and
semi-major axis of both MOs is shown in Fig.~\ref{pertMOs}a,b.
Specifically, the orbits of the MOs change significantly for a short
period of time. The distance between both MOs suddenly increases and
the mutual perturbation becomes insignificant. Then the original orbits
are almost restored.

% FIG. 9 (pertMOs)

   In addition, we also investigated the perturbation effect of the
second MO with mass $10^{-8}\,$M$_{\star}$ on the first MO, when the
mass of the latter is negligible, only $10^{-14}\,$M$_{\star}$. In this
case, the orbit of the second MO remained practically the same. As
expected, the orbit of the lighter MO changed more than in the case of
two equal-mass MOs. The change happened at the close approach of both
MOs, to a distance of only $0.00019\,$au, which occured $\sim$$8\,850$
days after the first-MO periastron passage, in this case. The change of
the periastron distance and semi-major axis of this MO is shown in
Fig.~\ref{pertMOs}c,d. Again, the orbital elements change their values
for only a short part of the duration of strong perturbation. Then the
original values are almost restored. Hence, our assumption of a series
of MOs moving in identical orbits is reasonable. We note, a new
cloud can be formed due to the tidal action of the perturbing MO at the
close approach, if the parent MO contains some volatile material.

\subsection{Perturbation on a tight dust cloud}

   In Sect.~\ref{6.1}, we considered the gravitational perturbation caused by
a second MO on the cloud of model (A) around the first MO. This cloud
appeared to be almost unaffected during the investigated period.
The DPs in this cloud are at relatively large distances from their
parent MO. Hence, the close approach of the second MO to the first MO
does not mean the close approach of the former to the DPs. This is,
however, no longer true if we consider a tight dust cloud, at a
relatively short distances from its parent MO. The DPs in the tight
cloud can be expected to be perturbed with almost the same strength as
the parent MO alone.

   To see the effect of the perturbation on the tight cloud we performed
the following two calculations. Firstly, we assumed a MO with
mass of $10^{-8}\,$M$_{\star}$ on the same orbit as in the first
simulation described in Sect.~\ref{twoMOs} and a tight dust cloud of
type (A) around it. The DPs in the cloud have $\beta = 0$ and
pericenters and apocenters of $1\,894\,$km (radius of the MO) and
$15\, 000\,$km, respectively, This is within the Hill radius of the MO,
which is approximately $22\, 000\,$km. Integration started at $r_{o}=5\,$au. In
the other simulation, we added a second MO with the same mass as the first. This
MO was released into the same orbit following the first MO after a time
delay of 20 days.

   The results of both simulations were compared. In the first
simulation, the cloud remains almost untouched after the single orbital
revolution. If the second MO is considered, the cloud survives until
the close mutual approach of both MOs. Then the pericenters of $16.6\%$
of the DPs are reduced and these DPs end up on the MO's surface. The
other DPs are detached from their parent MO and follow their own
trajectory around the central star. One might anticipate that during
such a bombardment of the first MO's surface, a sub-surface layer of
volatile material would be exposed, which could trigger an enhanced
activity upon the next approach to the star, in analogy to the activity
of Main Belt Comets \citep{haghighipour16}.

\subsection{Differences in the argument of periastron}

   Another problem to discuss is the difference in the angle $\Phi$ for
various features. The features at 800, $1\,520$, and $1\,540$ days
could all be fitted by models when value $\Phi = 29^{o}$ is considered.
The last feature, at $1\,570$ days, can be fitted with a significantly
lower value of $\Phi = 0^{o}$. Nevertheless, even such a relatively
large difference in the argument of periastron $\sim$$30^{o}$ in
$\omega$ might still allow a common progenitor. We can look for an
explanation in our own Solar System again, with the
groups of sun-grazing comets representing one particular inspiration. Namely, \cite{ohtsuka03} and
\cite{sekanina05} argued for a common origin, in the same progenitor, of
all Marsden and Kracht groups of sun-grazing comets, comet 96P/Machholz,
and the daytime Arietid meteoroid stream. Sekanina and Chodas found the
difference between the mean $\omega$ of Marsden and Kracht groups to be
$36.7^{o}$. \cite{ohtsuka03} presented the orbits of these groups with
an even larger maximum $\omega$-difference, equal to approximately $47^{o}$.

\subsection{Feature at 1$\,$210 days and miscellaneous comments}

\begin{figure}
\centerline{\includegraphics[height=8.5cm, angle=-90]{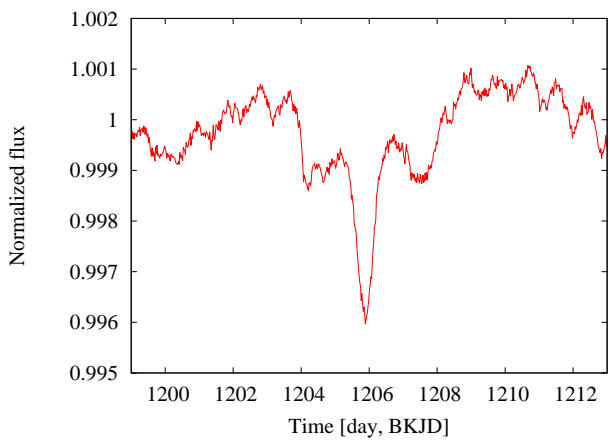}}
\caption{Tiny eclipse event observed near to day $1210$, which is very 
similar to the one observed at 1$\,$540 days, may have important consequences for
the different scenarios.}
\label{f1210}
\end{figure}

   We would like to point out that in this paper, we modeled only the four
strongest features observed in the Kepler light curve of this object.
Aside from these, there are several other, considerably fainter features. One of them is found at approximately 1$\,$210 days and deserves further attention. It is illustrated in Figure \ref{f1210}. This is a
symmetric triple peak structure with the central peak being the
strongest, and closely resembles the feature seen at 1$\,$540 days investigated as part of this study. The ratio
of the central to side peaks is almost the same but is, in fact, slightly
narrower. This similarity allows us to argue that it is also caused by
a similar object with a dusty ring, model (C).

   While this feature does not pose a significant problem for our model 
(it takes only one additional massive object with a dusty cloud on 
the same orbit), its existence renders many other theories much less 
plausible. For example, the comet scenario \citep{bodman16} would require
the comets to gather by chance into the same constellation as during the
1$\,$540 feature. Within the interstellar cloud, ISM structure, and a
dark disk with a black hole scenario \citep{wright16b} an accidental 
repetition of the same structure within the cloud would also be required.
%Within the free-floating interstellar object scenario...

   There are other tiny features observed at 140 and 260 days. These
features may show a `pre-transit' and `post-transit' brightening. Such
a brightening is most probably caused by the forward scattering of light
from the host star by the dust cloud. This would require that the cloud
be close to the star, that is, of circumstellar origin, and would rule out
all theories in which the eclipsing object is in the interstellar medium
or Solar System.

   An increasing fading of the star by approximately 3\% during the Kepler
mission may not be a problem for our model. Dust clouds or debris
associated with four massive objects may naturally extend and spread
along their orbit. The highest concentration should be near the objects
which is where it is observed.

% FIG. 10

\section{Conclusions}

   Our main findings and arguments are briefly summarized below.

\begin{enumerate}
\item We demonstrate that it is possible to explain the complex
morphology of the Kepler light-curve of KIC8462852 with a very simple
model. Only four massive objects, each surrounded by a dust cloud, can
account for most of the observed features. The objects are
apparently of a common origin, that is, the result of a break-up process of a
single progenitor.
\item Most of the features may be represented by a simple,
initially spherical dust cloud. Such clouds in eccentric orbits are
observed to naturally vertically shrink and develop a leading tail as
they approach periastron. The feature at 1$\,$540 days seems to be
special since it is best reproduced by an initially ring-like
structure.
\item This scenario of four massive objects is further supported by the
following arguments: the smooth shape of the 800-day feature, which
is difficult to assemble from a number of smaller objects such as
comets; a tendency towards shallower ingress and steeper egress of the
800-day feature, which is exactly the opposite of what is expected for the
less massive objects such as comets; the $1\,520$ and $1\,570$-day
features also show a gradual increase in strengths of individual
`sub-features' and fast recovery, resembling the 800-day feature;
the symmetric `ring-like' structure of the $1\,540$-day feature, which
would presume a non-negligible gravity of the object; the existence
of another symmetric structure at $1\,210$ days, which is very similar to
the above mentioned feature and is difficult to understand within
a comet scenario or other models; the clustering of the obscuration
features into four main events, which naturally leads to the association
with four objects; as well as by the fact that our solution indicates
that all four bodies are on very similar eccentric orbits. Further, all best fits were for the P-R drag parameter $\beta = 0.629,$ which
indicates that also the dust particles may be similar in size and
chemical composition.
\item It is not claimed that we found the only/best solution within this
concept of four massive bodies. We rather state that we found a possible
solution.
\item Iron or carbon grains smaller than approximately 0.1 microns experience a
very strong radiative push, which quickly disconnects them from the
parent body and places them on hyperbolic orbits. Thus, it is unlikely
that such grains contribute significantly to the observed features.
\item Grains larger than approximately 100 microns experience small radiative
accelerations and may remain bound to the massive object. Their opacity
is small, therefore they are not likely to contribute significantly to the
obscuration events. However, they may act as a reservoir and produce
smaller dust grains via collisions.
\item It is argued that 0.3$-$10 micron-sized dust grains are the best
candidates for explaining the obscuration events. Smaller grains,
unless they were being replenished, would be easily expelled from
the system while larger grains would have a relatively small opacity.
\item It was shown that the mutual interaction between the massive
objects and their dust clouds within few astronomical units from the
periastron can be neglected and that they can be treated independently
of each other in this region.
\item If the two massive objects follow each other on identical
eccentric orbits with a short enough time lag, a strong interaction 
between them and their dust clouds may happen at larger distances from
the star, which might disperse the clouds but, at the same time, also
expose sub-surface volatile material, trigger outbursts, and produce
debris.
\end{enumerate}

   The outlined concept provides an alternative explanation of
the observed light-curve of KIC8462852. It is a simple model with only
a small number of free parameters. Although the similarity between the observed and
simulated light-curves is striking, the fits can certainly be improved;
motivation for further research in this direction. There is a
plethora of possibilities of how to improve, modify, or advance such
models. Especially, more extensive calculations with more realistic dust
clouds combined with some minimization techniques might shed more light
onto what has happened around this interesting star. On the other hand,
at the moment, there is no need to invoke alien mega-structures to explain the above mentioned light-curves.

\begin{acknowledgements}
We thank David Vokrouhlick\'{y} and Oleksandra Ivanova for 
stimulating discussion and an anonymous referee for careful reading of 
the manuscript and comments that helped to improve the paper 
considerably.
This work was supported by VEGA - the Slovak Grant Agency for Science, 
grants Nos. 2/0143/14 and 2/0031/14 as well as by the APVV 15-0458
project. This article was also created by the realization of the project
ITMS No. 26220120029, based on the supporting operational Research and
development program financed by the European Regional Development Fund.
\end{acknowledgements}

\bibliographystyle{aa} % style aa.bst
\bibliography{budaj,neslusan} % your references Yourfile.bib

\begin{appendix}

\section{Characteristics of all models}

% TAB. A.1
\begin{table*}
\caption{The initial characteristics of A-type models with 1$\,$000 test
particles and the massive object in `standard orbit' (periastron
distance of $0.1\,$au; apastron distance of $50\,$au) as well as in the
standard initial star-centric distance ($5\,$au). Symbols used: $M_{mo}$
$-$ mass of the `MO, $q_{tp}$ and $Q_{tp}$ $-$ pericenter and
apocenter distance of the test particles with respect to the MO, and $\beta$ $-$ parameter characterizing the strength of
the Poynting-Robertson drag. In models A59$-$A62, MO-centric
velocity of the particles in the pericenter, $v_{q}$, is given (in
m$\,$s$^{-1}$) instead of $Q_{tp}$.}
% N = 1000; q = 0.1 AU; Q = 50 AU; r_{o} = 5 AU
\begin{tabular}{rcccl}
\hline \hline
No. & $M_{mo}$ & $q_{tp}$ & $Q_{tp}$ & $\beta$ \\ % subdir.
    & $[$M$_{\star}$$]$ & $[$km$]$ & $[$km$]$ & $[$1$]$ \\
\hline
 A1 & $10^{-10}$ & $10^{3}$ & $10^{5}$ &  0.007 \\
 A2 &            &          &          & 0.07 \\
 A3 &            &          &          & 0.35 \\
 A4 &            &          &          & 0.629 \\ % GRID1m1m10B*
 A5 & $10^{-12}$ & $10^{3}$ & $10^{5}$ &  0.007 \\
 A6 &            &          &          & 0.07 \\
 A7 &            &          &          & 0.35 \\
 A8 &            &          &          & 0.629 \\ % GRID1m1m12B*
 A9 & $10^{-13}$ & $10^{3}$ & $10^{5}$ &  0.007 \\
A10 &            &          &          & 0.07 \\
A11 &            &          &          & 0.35 \\
A12 &            &          &          & 0.629 \\ % GRID1m1m13B*
A13 & $10^{-14}$ & $10^{3}$ & $10^{5}$ &  0.007 \\
A14 &            &          &          & 0.07 \\
A15 &            &          &          & 0.35 \\
A16 &            &          &          & 0.629 \\ % GRID1m1m14B*
A17 & $10^{-8}$  & $10^{3}$ & $10^{5}$ &  0.000007 \\
A18 &            &          &          & 0.007 \\
A19 &            &          &          & 0.07 \\
A20 &            &          &          & 0.35 \\
A21 &            &          &          & 0.629 \\ % GRID1m1m8B*
A22 & $10^{-11}$ &    100   &   100    &  0.629 \\ % q01qtp100Qtp1m11B0.9
A23 & $10^{-11}$ &    100   & $10^{3}$ &  0.629 \\ % q01qtp100Qtp10m11B0.9
A24 & $10^{-12}$ &    100   & $10^{4}$ &  0.629 \\ % q01qtp100Qtp100m12B0.9
A25 & $10^{-8}$  &    100   & $10^{5}$ &  0.629 \\ % q01qtp100Qtp1000m8B0.9
A26 & $10^{-9}$  &    100   & $10^{5}$ &  0.629 \\ % q01qtp100Qtp1000m9B0.9
A27 & $10^{-10}$ &    100   & $10^{5}$ &  0.629 \\ % q01qtp100Qtp1000m10B0.9
A28 & $10^{-11}$ &    100   & $10^{5}$ &  0.629 \\ % q01qtp100Qtp1000m11B0.9
A29 & $10^{-12}$ &    100   & $10^{5}$ &  0.629 \\ % q01qtp100Qtp1000m12B0.9
A30 & $10^{-11}$ &    100   &   500    &  0.629 \\ % q01qtp100Qtp5m11B0.9
A31 & $10^{-11}$ &    100   & $5\times 10^{4}$ &  0.629 \\ % q01qtp100Qtp500m11B0.9
A32 & $10^{-9}$  &     50   & $10^{5}$ & 0.629 \\ % q01qtp50Qtp2000m9B0.9
A33 & $10^{-8}$  &     75   & $7.5\times 10^{4}$ & 0.8 \\ % GRID1m1m8B1.144
A34 & $10^{-8}$  &          &                    & 0.9 \\ % GRID1m1m8B1.287
A35 & $10^{-9}$  &     75   & $7.5\times 10^{4}$ & 0.8 \\ % GRID1m1m9B1.144
A36 & $10^{-9}$  &          &                    & 0.9 \\ % GRID1m1m9B1.287
A37 & $10^{-10}$ &     75   & $7.5\times 10^{4}$ & 0.8 \\ % GRID1m1m10B1.144
A38 & $10^{-10}$ &          &                    & 0.9 \\ % GRID1m1m10B1.287
A39 & $10^{-11}$ &     75   & $7.5\times 10^{4}$ & 0.629 \\ % q01qtp75Qtp1000m11B0.9
A40 & $10^{-11}$ &          &                    & 0.8 \\ % GRID1m1m11B1.144
A41 & $10^{-11}$ &          &                    & 0.9 \\ % GRID1m1m11B1.287
A42 & $10^{-12}$ &     75   & $7.5\times 10^{4}$ & 0.8 \\ % GRID1m1m12B1.144
A43 & $10^{-12}$ &          &                    & 0.9 \\ % GRID1m1m12B1.287
A44 & $10^{-12}$ & $10^{3}$ & $10^{6}$ &  0.007 \\
A45 &            &          &          & 0.07 \\
A46 &            &          &          & 0.629 \\
A47 &            &          &          & 1.4 \\ % q01Q1e3qm1m12B*
A48 & $10^{-10}$ & $10^{3}$ & $10^{4}$ &  0.35 \\
A49 &            &          &          & 0.629 \\ % q01Q10qm1m10B*
A50 & $10^{-10}$ & $10^{3}$ & $10^{5}$ & 1.4 \\ % q01Q100qm1m10B2
A51 & $10^{-12}$ & $10^{3}$ & $10^{5}$ & 1.4 \\ % q01Q100qm1m12B2
A52 & $10^{-7}$  & $10^{3}$ & $10^{5}$ &  0.007 \\
A53 &            &          &          & 0.07 \\
A54 &            &          &          & 0.629 \\
A55 &            &          &          & 1.4 \\ % q01Q100qm1m7B*
\hline
\end{tabular}
\end{table*}

\begin{table*}
\begin{tabular}{rcccl}
\hline
A56 & $10^{-8}$  & $10^{3}$ & $10^{5}$ & 1.4 \\ % q01Q100qm1m8B2
A57 & $10^{-10}$ & $10^{3}$ & $1.1\times 10^{3}$ & 0.629 \\ % q01Q1.1qm1m10B0.9
A58 & $10^{-10}$ & $10^{3}$ & $3\times 10^{3}$ & 0.829 \\ % q01Q3qm1m10B0.9
A59 & $10^{-12}$ & $10^{3}$ & $v_{q}=300$ &  0.007 \\
A60 &            &          &          & 0.07 \\
A61 &            &          &          & 0.629 \\
A62 &            &          &          & 1.4 \\ % comv300B*
\hline \hline
\end{tabular}
\end{table*}

% TAB. A.2
\begin{table*}
\caption{The initial characteristics of A-type models with the number
of test particles, $N$, other than 1$\,$000 and the massive object in
`standard orbit' (periastron distance of $0.1\,$au; apastron distance of
$50\,$au) as well as in the standard initial star-centric distance
($5\,$au). The same symbols as in Table~A.1 are used.}
% N \neq 1000; q = 0.1 AU; Q = 50 AU; r_{o} = 5 AU
\begin{tabular}{rccccl}
\hline \hline
No. & $N$ & $M_{mo}$ & $q_{tp}$ & $Q_{tp}$ & $\beta$ \\ % subdir.
  & $[$1$]$ & $[$M$_{\star}$$]$ & $[$km$]$ & $[$km$]$ & $[$1$]$ \\
\hline
A63 & $2\times 10^{3}$ & $10^{-12}$ & $10^{3}$ & $v_{q}=300$ & 0.0175 \\
A64 &          &            &          &          & 0.07 \\ % Cq01r5B*
A65 & $10^{4}$ & $10^{-10}$ & $10^{3}$ & $10^{5}$ & 0.28 \\
A66 &          &            &          &          & 0.35 \\
A67 &          &            &          &          & 0.42 \\
A68 &          &            &          &          & 0.49 \\ % GRID1m1m10B*
A69 & $10^{4}$ & $10^{-8}$  & $10^{3}$ & $10^{5}$ & 0.007 \\
A70 &          &            &          &          & 0.07 \\
A71 &          &            &          &          & 0.629 \\ % q01m1m8B*
A72 & $10^{4}$ & $10^{-9}$  & $10^{3}$ & $10^{5}$ & 0.28 \\
A73 &          &            &          &          & 0.35 \\
A74 &          &            &          &          & 0.42 \\
A75 &          &            &          &          & 0.49 \\ % GRID1m1m9B*
A76 & $10^{4}$ & $10^{-11}$ & $10^{3}$ & $10^{5}$ & 0.28 \\
A77 &          &            &          &          & 0.35 \\
A78 &          &            &          &          & 0.42 \\
A79 &          &            &          &          & 0.49 \\ % GRID1m1m11B*
A80 & $2\times 10^{3}$ & $10^{-12}$ & $10^{6}$ & $v_{q}=300$ & 0.0175 \\
A81 &          &            &          &          & 0.07 \\ % Cq01r5B*
\hline \hline
\end{tabular}
\end{table*}

% TAB. A.3
\begin{table*}
\caption{The initial characteristics of A-type models with the massive
object in orbits other than the `standard orbit' (periastron distance of $0.1\,$au;
apastron distance of $50\,$au) and/or at distances other than the standard initial
star-centric distance ($5\,$au). The periastron and apastron of the
massive-object orbit are denoted by $q$ and $Q$. The initial star-centric
distance of this object is denoted by $r_{o}$ and the number of test
particles by $N$. The other symbols are the same as in Table~A.1. $q$ and
$r_{o}$ are given either in the radii of the central star, R$_{\star}$, or
in astronomical units. The velocity in the pericenter, $v_{q}$, with
respect to the massive object is given as a multiple of escape velocity
from this object, v$_{ii}$.}
% the other A models
\begin{tabular}{rcccccccl}
\hline \hline
No. & $q$ & $Q$ & $r_{o}$ & $N$ & $M_{mo}$ & $q_{tp}$ & $Q_{tp}$ &
    $\beta$ \\ % subdir.
    &   & $[$au$]$ &     & $[$1$]$ & $[$M$_{\star}$$]$ & $[$km$]$ &
  $[$km$]$ & $[$1$]$ \\
\hline
 A82 & $1.4\,$R$_{\star}$ & 40 & $1.6\,$R$_{\star}$ & $10^{3}$ &
 $10^{-7}$ & $10^{4}$ & $10^{5}$ & 0.007 \\
 A83 &               &    &               &        &          &
          &  &  0.07 \\
 A84 &               &    &               &        &          &
          &  &  0.629 \\
 A85 &               &    &               &        &          &
          &  & 1.4 \\ % EQ10qm1m7B*
 A86 & $1.5\,$R$_{\star}$ & 50 &   $5\,$au & $10^{3}$ &
 $10^{-7}$ & $10^{3}$ & $10^{5}$ & 0.007 \\
 A87 &               &    &               &        &          &
          &  &  0.07 \\
 A88 &               &    &               &        &          &
          &  &  0.629 \\
 A89 &               &    &               &        &          &
          &  & 1.4 \\ % FQ100qm1m7B*
 A90 & $1.5\,$R$_{\star}$ & 50 &   $5\,$au & $10^{3}$ &
 $10^{-9}$ & $10^{3}$ & $10^{5}$ & 0.007 \\
 A91 &               &    &               &        &          &
          &  &  0.07 \\
 A92 &               &    &               &        &          &
          &  &  0.629 \\
 A93 &               &    &               &        &          &
          &  & 1.4 \\ % FQ100qm1m9B*
 A94 & $1.5\,$R$_{\star}$ & 50 &   $1\,$au & $10^{3}$ &
 $10^{-7}$ & $10^{3}$ & $10^{4}$ & 0.007 \\
 A95 &               &    &               &        &          &
          &  &  0.07 \\
 A96 &               &    &               &        &          &
          &  & 0.629 \\
 A97 &               &    &               &        &          &
          &  & 0.769 \\
 A98 &               &    &               &        &          &
          &  & 1.4 \\ % IQ10qm1m7B*
 A99 & $1.5\,$R$_{\star}$ & 50 &   $1\,$au & $10^{3}$ &
 $10^{-7}$ & $10^{3}$ & $10^{5}$ & 0.007 \\
 A100 &               &    &               &        &          &
          &  &  0.07 \\
 A101 &               &    &               &        &          &
          &  & 0.629 \\
 A102 &               &    &               &        &          &
          &  & 1.4 \\ % IQ10qm1m7B*
 A103 & $0.1\,$au     & 50 &   $3\,$au     & $10^{3}$ &
 $10^{-10}$ & $10^{3}$ & $10^{5}$ & 0.629 \\ % Jq01Q100qm1m10B0.9
 A104 & $1.5\,$R$_{\star}$ & 50 &   $1\,$au & $10^{3}$ &
 $10^{-7}$  & $10^{4}$ & $10^{5}$ & 0.007 \\
 A105 &               &    &               &        &          &
          &  &  0.07 \\
 A106 &               &    &               &        &          &
          &  & 0.629 \\
 A107 &               &    &               &        &          &
          &  & 0.769 \\
 A108 &               &    &               &        &          &
          &  & 1.4 \\ % JQ10qm1m7B*
 A109 & $1.5\,$R$_{\star}$ & 50 & $1.5\,$R$_{\star}$ & $10^{3}$ &
 $10^{-7}$  & $10^{4}$ & $10^{5}$ & 0.007 \\
A110 &               &    &               &        &          &
          &  & 0.07 \\
A111 &               &    &               &        &          &
          &  & 0.629 \\
A112 &               &    &               &        &          &
          &  & 1.4 \\ % PQ10qm1m7B*
A113 & $0.05\,$au    & 50 &   $5\,$au     & $10^{3}$ &
 $10^{-10}$ & $10^{3}$ & $10^{4}$ & 0.629 \\ %  q005Q10qm1m10B0.9
A114 & $0.05\,$au    & 50 &   $5\,$au     & $10^{3}$ &
 $10^{-10}$ & $10^{3}$ & $10^{5}$ & 0.629 \\ % q005Q100qm1m10B0.9
A115 & $0.1\,$au     & 50 &  $10\,$au     & $10^{3}$ &
 $10^{-11}$ & 75 & $7.5\times 10^{4}$ & 0.629 \\ % q01qtp75Qto1000m11B0.9ro10
A116 & $0.1\,$au     & 50 &  $10\,$au     & $10^{3}$ &
 $10^{-12}$ & 75 & $7.5\times 10^{4}$ & 0.629 \\ % q01qtp75Qto1000m12B0.9ro10
A117 & $0.2\,$au & 15 & $5\,$au  & $10^{3}$ &
 $10^{-7}$  & $10^{3}$ & $10^{5}$ & 0.629 \\ % q02Q15m1m7B0.9
A118 & $0.2\,$au & 15 & $5\,$au & $10^{3}$ &
 $10^{-10}$ & $10^{3}$ & $10^{5}$ & 0.007 \\
A119 &               &    &               &        &          &
          &  & 0.629 \\ % q02Q15m1m10B*
A120 &   $1\,$au & 50 & $5\,$au & $10^{3}$ &
 $10^{-10}$ & $10^{3}$ & $10^{5}$ & 0.07 \\
A121 &                    &               &        &          &
          &  &  & 0.629 \\ % q1Q100qm1m10B*
A122 & $0.1\,$au & 50 & $3\,$au & $10^{3}$ &
 $10^{-10}$ & $10^{3}$ & $10^{4}$ & 0.629 \\ % R3q01Q10qm1m10B0.9
A123 & $0.1\,$au & 50 & $7\,$au & $10^{3}$ &
 $10^{-10}$ & $10^{3}$ & $10^{4}$ & 0.629 \\ % R7q01Q10qm1m10B0.9
A124 & $0.1\,$au & 50 & $10\,$au & $10^{3}$ &
 $10^{-10}$ & $10^{3}$ & $10^{4}$ & 0.629 \\ %  R10q01Q10qm1m10B0.9
A125 & $1.5\,$R$_{\star}$ & 50 & $5\,$au & $10^{4}$ &
 $10^{-7}$  & $3\,238$ & $6\,476$ & 0.07 \\
A126 &               &    &               &        &          &
          &   & 0.35 \\
A127 &               &    &               &        &          &
          &   & 0.629 \\ % Bq1.5m1m7B*
A128 & $1.5\,$R$_{\star}$ & 50 & $5\,$au & $10^{4}$ &
 $10^{-8}$  & $1\,503$ & $3\,006$ &   0.007 \\
A129 &               &    &               &        &          &
          &  & 0.07 \\
A130 &               &    &               &        &          &
          &  & 0.35 \\
A131 &               &    &               &        &          &
          &  & 0.629 \\ % Bq1.5m1m8B*
A132 & $1.4\,$R$_{\star}$ & 40 & $1.6\,$R$_{\star}$ & $10^{3}$ &
 $10^{-7}$  & $10^{4}$ & $v_{q}=0.1\,$v$_{ii}$ & 0.007 \\
A133 &               &    &               &        &          &
          &  & 0.07 \\
A134 &               &    &               &        &          &
          &  & 0.629 \\ % Ev0.1m1m7B*
A135 & $1.4\,$R$_{\star}$ & 40 & $1.6\,$R$_{\star}$ & $10^{3}$ &
 $10^{-9}$  & $10^{3}$ & $v_{q}=0.1\,$v$_{ii}$ & 0.007 \\
\hline
\end{tabular}
\end{table*}

% TAB. A.3 - part 2
\begin{table*}
\begin{tabular}{rcccccccl}
\hline
A136 &               &    &               &        &          &
          &  & 0.07 \\
A137 &               &    &               &        &          &
          &  & 0.629 \\ % Ev0.1m1m9B*
A138 & $1.4\,$R$_{\star}$ & 40 & $1.6\,$R$_{\star}$ & $10^{3}$ &
 $10^{-9}$  & $10^{3}$ & $v_{q}=\,$v$_{ii}$ & 0.007 \\
A139 &               &    &               &        &          &
          &  & 0.07 \\
A140 &               &    &               &        &          &
          &  & 0.629 \\ % Ev1m1m9B*
A141 & $1.5\,$R$_{\star}$ & 50 & $1\,$au & $10^{3}$ &
 $10^{-7}$  & $10^{4}$ & $v_{q}=0.1\,$v$_{ii}$ & 0.007 \\
A142 &               &    &               &        &          &
          &  & 0.07 \\
A143 &               &    &               &        &          &
          &  & 0.629 \\ % Jv0.1m1m7B*
A144 & $1.5\,$R$_{\star}$ & 50 & $1.5\,$R$_{\star}$ & $10^{3}$ &
 $10^{-7}$  & $10^{4}$ & $v_{q}=0.1\,$v$_{ii}$ & 0.01 \\
A145 &               &    &               &        &          &
          &  & 0.07 \\
A146 &               &    &               &        &          &
          &  & 0.629 \\ % Pv0.1m1m7B*
A147 & $1.5\,$R$_{\star}$ & 50 & $1.5\,$R$_{\star}$ & $10^{3}$ &
 $10^{-9}$  & $10^{4}$ & $v_{q}=0.1\,$v$_{ii}$ & 0.007 \\
A148 &               &    &               &        &          &
          &  &  0.07 \\
A149 &               &    &               &        &          &
          &  & 0.629 \\ % Pv0.1m1m9B*
A150 & $1.5\,$R$_{\star}$ & 50 & $1.5\,$R$_{\star}$ & $10^{3}$ &
 $10^{-7}$  & $10^{4}$ & $v_{q}=\,$v$_{ii}$ & 0.007 \\
A151 &               &    &               &        &          &
          &  & 0.07 \\
A152 &               &    &               &        &          &
          &  & 0.629 \\ % Pv1m1m7B*
A153 & $1.5\,$R$_{\star}$ & 50 & $1.5\,$R$_{\star}$ & $10^{3}$ &
 $10^{-9}$  & $10^{4}$ & $v_{q}=\,$v$_{ii}$ & 0.007 \\
A154 &               &    &               &        &          &
          &  & 0.07 \\
A155 &               &    &               &        &          &
          &  & 0.629 \\ % Pv1m1m9B*
A156 & $1.5\,$R$_{\star}$ & 50 & $1.5\,$R$_{\star}$ & $10^{3}$ &
 $10^{-7}$  & $10^{4}$ & $v_{q}=10\,$v$_{ii}$ & 0.007 \\
A157 &               &    &               &        &          &
          &  & 0.07 \\
A158 &               &    &               &        &          &
          &  & 0.629 \\ % Pv10m1m7B*
A159 & $10\,$R$_{\star}$ & 50 & $5\,$au & $10^{4}$ &
 $10^{-12}$ & 100 & $10^{3}$ &  0.07 \\ % q10Q50m1m12B0.1
A160 & $1.5\,$R$_{\star}$ & 50 & $5\,$au & $10^{4}$ &
 $10^{-7}$  & $10^{3}$ & $10^{5}$ &  0.07 \\
A161 &               &    &               &        &          &
          &  & 0.35 \\
A162 &               &    &               &        &          &
          &  & 0.629 \\ % q1.5m1m7B*
A163 & $1.5\,$R$_{\star}$ & 50 & $5\,$au & $10^{4}$ &
 $10^{-8}$  & $10^{3}$ & $10^{5}$ &  0.007 \\
A164 &               &    &               &        &          &
          &  & 0.07 \\
A165 &               &    &               &        &          &
          &  & 0.629 \\ % q1.5m1m8B*
A166 & $1.5\,$R$_{\star}$ & 50 & $5\,$au & $10^{4}$ &
 $10^{-10}$ & $10^{3}$ & $10^{5}$ &  0.07 \\
A167 &               &    &               &        &          &
          &  & 0.35 \\
A168 &               &    &               &        &          &
          &  & 0.629 \\ % q1.5m1m10B*
A169 & $2\,$R$_{\star}$ & 50 & $5\,$au & $10^{4}$ &
 $10^{-8}$  & $10^{3}$ & $10^{5}$ &  0.07 \\ % q2Q50m1m8v100B0.1
A170 & $4\,$R$_{\star}$ & 50 & $5\,$au & $10^{4}$ &
 $10^{-12}$ & 100 & $10^{3}$ &  0.07 \\
A171 &               &    &               &        &          &
          &  & 0.629 \\ % q4Q50m1m12B*
A172 & $1.5\,$R$_{\star}$ & 50 & $5\,$au & $10^{3}$ &
 $10^{-7}$  & $5\times 10^{3}$ & $10^{4}$ & 0.007 \\
A173 &               &    &               &        &          &
          &  & 0.07 \\
A174 &               &    &               &        &          &
          &  & 0.629 \\ % TBq1.5m1m7B*
A175 & $1.5\,$R$_{\star}$ & 50 & $5\,$au & $10^{3}$ &
 $10^{-10}$ & $5\times 10^{3}$ & $10^{4}$ & 0.007 \\
A176 &               &    &               &        &          &
          &  & 0.07 \\
A177 &               &    &               &        &          &
          &  & 0.629 \\ % TBq1.5m1m10B*
A178 & $1.5\,$R$_{\star}$ & 50 & $5\,$au & $10^{3}$ &
 $10^{-9}$  & $10^{3}$ & $10^{4}$ &  0.007 \\
A179 &               &    &               &        &          &
          &  & 0.07 \\
A180 &               &    &               &        &          &
          &  & 0.629 \\ % Tq1.5m1m9B*
A181 &  $0.1\,$au & 50 & $3\,$au & $10^{3}$ &
 $10^{-10}$ & $10^{3}$ & $10^{5}$ &  0.629 \\ % Jq01Q100qm1m10B0.9
A182 &  $0.05\,$au & 50 & $5\,$au & $10^{3}$ &
 $10^{-10}$ & $10^{3}$ & $10^{5}$ &  0.629 \\ % q005Q100qm1m10B0.9
A183 &  $0.05\,$au & 50 & $5\,$au & $10^{3}$ &
 $10^{-10}$ & $10^{3}$ & $10^{4}$ &  0.629 \\ % q005Q10qm1m10B0.9
A184 & $10\,$R$_{\star}$ & 50 & $5\,$au & $10^{4}$ &
 $10^{-12}$ & 100 & $10^{3}$ &  0.07 \\ % q10Q50m1m12B0.1
A185 &  $0.1\,$au & 15 & $5\,$au & $10^{3}$ &
 $10^{-10}$ & $10^{3}$ & $10^{5}$ &  0.07 \\
A186 &               &    &               &        &          &
          &  & 0.35 \\
A187 &               &    &               &        &          &
          &  & 0.629 \\ % q01Q15m1m10B*
\hline \hline
\end{tabular}
\end{table*}

% TAB. A.4
\begin{table*}
\caption{The initial characteristics of B-type models. In all these
models, we consider $10^{3}$ test particles and the massive object
moving in the standard orbit having the periastron equal to $0.1\,$au,
apastron $50\,$au, and initial star-centric distance $5\,$au.
$Q_{tp;min}$ and $Q_{tp;max}$ are the minimum and maximum apocenter
distances of test particles orbiting the massive object in a cloud.
The other denotations are the same as in Tables~A.1 and A.3.}
\begin{tabular}{rccccl}
\hline \hline 
No. & $M_{mo}$ & $q_{tp}$ & $Q_{tp;min}$ & $Q_{tp;max}$ & $\beta$ \\ % subdir.
   & $[$M$_{\star}$$]$ & $[$km$]$ & $[$km$]$ & $[$km$]$ & $[$1$]$ \\
\hline
 B1 & $10^{-8}$  &  500   & $5\times 10^{3}$ & $5\times 10^{4}$ &
 0.629 \\ % GRIDn2m1m8B0.9q05
 B2 & $10^{-10}$ &  500   & $5\times 10^{3}$ & $5\times 10^{4}$ &
 0.629 \\ % GRIDn2m1m10B0.9q05
 B3 & $10^{-12}$ &  500   & $5\times 10^{3}$ &  $5\times 10^{4}$ &
 0.629 \\ % GRIDn2m1m12B0.9q05
 B4 & $10^{-8}$  & $10^{3}$ & $10^{5}$ & $10^{6}$ &
 0.629 \\ % GRIDn2m1m8B0.9a10
 B5 & $10^{-8}$  & $10^{3}$ & $3\times 10^{4}$ & $3\times 10^{5}$ &
 0.629 \\ % GRIDn2m1m8B0.9a3
 B6 & $10^{-10}$ & $10^{3}$ & $3\times 10^{4}$ & $3\times 10^{5}$ &
 0.629 \\ % GRIDn2m1m10B0.9a3
 B7 & $10^{-12}$ & $10^{3}$ & $10^{5}$ & $10^{6}$ &
 0.629 \\ % GRIDn2m1m12B0.9a10
 B8 & $10^{-12}$ & $10^{3}$ & $3\times 10^{4}$ & $3\times 10^{5}$ &
 0.629 \\ % GRIDn2m1m12B0.9a3
 B9 & $10^{-8}$  & $10^{3}$ & $10^{4}$ & $10^{5}$ & 0.07 \\
B10 &            &        &        &           & 0.35 \\
B11 &            &        &        &           & 0.629 \\ % GRID2m1m8B*
B12 & $10^{-8}$  & $10^{3}$ & $10^{4}$ & $10^{5}$ & 0.8 \\ % GRID2m1m8B1.144 \\
B13 &            &          &          &          & 0.9 \\ % GRID2m1m8B1.287 \\
B14 & $10^{-9}$  & $10^{3}$ & $10^{4}$ & $10^{5}$ & 0.8 \\ % GRID2m1m9B1.144 \\
B15 &            &          &          &          & 0.9 \\ % GRID2m1m9B1.287 \\
B16 & $10^{-10}$ & $10^{3}$ & $10^{4}$ & $10^{5}$ & 0.07 \\
B17 &            &        &        &              & 0.35 \\
B18 &            &        &        &              & 0.629 \\
B19 &            &        &        &              & 0.8 \\
B20 &            &        &        &              & 0.9 \\ % GRID2m1m10B*
B21 & $10^{-11}$ & $10^{3}$ & $10^{4}$ & $10^{5}$ & 0.8 \\ % GRID2m1m11B1.144 \\
B22 &            &          &          &          & 0.9 \\ % GRID2m1m11B1.287 \\
B23 & $10^{-12}$ & $10^{3}$ & $10^{4}$ & $10^{5}$ & 0.07 \\
B24 &            &        &        &           & 0.35 \\
B25 &            &        &        &           & 0.629 \\
B26 &            &        &        &           & 0.8 \\
B27 &            &        &        &           & 0.9 \\ % GRID2m1m12B*
\hline \hline
\end{tabular}
\end{table*}

% TAB. A.5
\begin{table*}
\caption{The initial characteristics of C-type models. In all these
models, we consider $10^{3}$ test particles and the massive object
moving in the standard orbit having the periastron equal to $0.1\,$au
and apastron $50\,$au. $\vartheta$ and $\sigma$ are the angles
characterizing the orientation of the ring. $R_{min}$ and $R_{max}$ are
the radii of its inner and outer border. The other denotations are the
same as in Tables~A.1 and A.3. Remark$^{*}$: in models C36 to C39,
the distribution of the DPs in the ring is not uniform, but their radial
profile is generated using the formula for the MO-centric distance of
$j$-th DP $r_{j} = 7\,500 \pm 2\,500 \eta^{2}$
($r_{j} = 7\,500 \pm 2\,500 \eta^{3/2}$) in kilometers. The sign in
pair $\pm$ is randomly generated and $\eta$ is a random number from
the interval $(0,~1)$.}
\begin{tabular}{rccccccl}
\hline \hline
No. & $r_{o}$ & $M_{mo}$ & $\vartheta$ & $\sigma$ & $R_{min}$ & $R_{max}$
 & $\beta$ \\ % subdir.
  & $[$au$]$ & $[$M$_{\star}$$]$ & $[$deg$]$ & $[$deg$]$ & $[$km$]$ &
  $[$km$]$ & $[$1$]$ \\
\hline
 C1 & 5 & $10^{-9}$ &  30 &  45 &  879    & $1\,758$ & 0.07 \\
 C2 &   &        &      &     &         &       & 0.35 \\
 C3 &   &        &      &     &         &       & 0.629 \\ % GRID6Bm1m9B*
 C4 & 5 & $10^{-8}$ &  30 &  45 & $5\times 10^{3}$ & $10^{4}$ & 0.07 \\
 C5 &   &        &      &     &         &       & 0.35 \\
 C6 &   &        &      &     &         &       & 0.42 \\  % GRID6m1m8B*
 C7 & 7 & $10^{-8}$ &  30 &  45 & $5\times 10^{3}$ & $10^{4}$ & 0.629 \\
%                                                   GRID6m1m8B0.9ro7
 C8 & 5 & $10^{-9}$ &  30 &  45 & $5\times 10^{3}$ & $10^{4}$ & 0.07 \\
 C9 &   &        &      &     &         &       & 0.21 \\
C10 &   &        &      &     &         &       & 0.35 \\
C11 &   &        &      &     &         &       & 0.42 \\
C12 &   &        &      &     &         &       & 0.629 \\ % GRID6m1m9B*
C13 & 5 & $10^{-9}$ &  30 &  45 & $7\times 10^{3}$ &
 $1.4\times 10^{4}$ & 0.49 \\ % GRID6m1m9B0.7Cring
C14 & 5 & $10^{-9}$ &  30 &  45 & $1.08\times 10^{4}$ &
 $2.1\times 10^{4}$ & 0.629 \\ % GRID6m1m9B0.9Bring
C15 & 5 & $10^{-8}$ & 210 & 330 & $5\times 10^{3}$ & $10^{4}$ & 0.629 \\
%                                            GRID6m1m8B0.9the210sig330
C16 & 5 & $10^{-8}$ & 210 &  45 & $5\times 10^{3}$ & $10^{4}$ & 0.629 \\
%                                             GRID6m1m8B0.9the210sig45
C17 & 5 & $10^{-8}$ & 225 & 315 & $5\times 10^{3}$ & $10^{4}$ & 0.629 \\
%                                            GRID6m1m8B0.9the225sig315
C18 & 5 & $10^{-8}$ &  30 & 225 & $5\times 10^{3}$ & $10^{4}$ & 0.629 \\
%                                             GRID6m1m8B0.9the30sig225
C19 & 5 & $10^{-8}$ &  30 & 315 & $5\times 10^{3}$ & $10^{4}$ & 0.629 \\
%                                             GRID6m1m8B0.9the30sig315
C20 & 5 & $10^{-8}$ & 330 &  45 & $5\times 10^{3}$ & $10^{4}$ & 0.629 \\
%                                             GRID6m1m8B0.9the330sig45
%%% PRIDANE:
C21 & 5 & $10^{-8}$  &  30 &  45 & $1\,503$ & $22\,347$ & 0.07 \\
C22 &   &            &     &     &          &           & 0.629 \\
C23 &   &            &     &     &   & & 1.4 \\ % GRID7m1m8B*the30sig45
C24 & 5 & $10^{-9}$  &  30 &  45 &    698   & $10\,373$ & 0.07 \\
C25 &   &            &     &     &          &           & 0.629 \\
C26 &   &            &     &     &   & & 1.4 \\ % GRID7m1m9B*the30sig45
C27 & 5 & $10^{-10}$ &  30 &  45 &    324   &  $4\,815$ & 0.07 \\
C28 &   &            &     &     &          &           & 0.629 \\
C29 &   &            &     &     &   & & 1.4 \\ % GRID7m1m10B*the30sig45
C30 & 5 & $10^{-11}$ &  30 &  45 &    150   &  $2\,235$ & 0.07 \\
C31 &   &            &     &     &          &           & 0.629 \\
C32 &   &            &     &     &   & & 1.4 \\ % GRID7m1m11B*the30sig45
C33 & 5 & $10^{-12}$ &  30 &  45 &     70   &  $1\,037$ & 0.07 \\
C34 &   &            &     &     &          &           & 0.629 \\
C35 &   &            &     &     &   & & 1.4 \\ % GRID7m1m12B*the30sig45
C36$^{*}$ & 5 & $10^{-8}$  &  30 &  45 & $5\times 10^{3}$ & $10^{4}$
                                     & 0.629 \\ % GRID9m1m8B0.9
C37$^{*}$ & 5 & $10^{-8}$  &  30 &  45 & $5\times 10^{3}$ & $10^{4}$
                                     & 0.629 \\ % GRID10m1m8B0.9
C38$^{*}$ &   &    &     &     &  &  & 0.8 \\ % GRID10m1m8B1.144
C39$^{*}$ &   &    &     &     &  &  & 0.9 \\ % GRID10m1m8B1.287
\hline \hline
\end{tabular}
\end{table*}

\end{appendix}

\end{document}